\documentclass{article}
\usepackage{amsmath}
\usepackage{amsfonts}
\usepackage{mathtools}
\usepackage{graphicx}
\usepackage{subfigure}
\usepackage{url}
\usepackage{color}
\usepackage{mathrsfs}
\usepackage{pifont}
\usepackage{natbib}
\usepackage{geometry}
\usepackage{fleqn}
\usepackage{txfonts}
\usepackage{hyperref}
\usepackage{natbib}
\usepackage{multirow}
\usepackage{abstract}

\newcommand{\defeq}{\vcentcolon=}
\usepackage[T1]{fontenc}
\usepackage[utf8]{inputenc}
\usepackage{authblk}
\title{Predictive modelling of a novel anti-adhesion therapy to combat bacterial colonisation of burn wounds}
\author[1,2]{Paul A. Roberts\footnote{Corresponding author\\E-mail address: p.a.roberts@univ.oxon.org (Paul A. Roberts)}}
\author[3]{Ryan M. Huebinger}
\author[2]{Emma Keen}
\author[4]{Anne-Marie Krachler}
\author[1,2]{Sara Jabbari}
\affil[1]{(School of Mathematics, University of Birmingham, Edgbaston Campus, Birmingham, UK)}
\affil[2]{(Institute of Microbiology and Infection, School of Biosciences, University of Birmingham, Edgbaston Campus, Birmingham, UK)}
\affil[3]{(Department of Surgery, University of Texas Southwestern Medical Center, Dallas, TX, USA)}
\affil[4]{(Department of Microbiology and Molecular Genetics, University of Texas McGovern Medical School at Houston, Houston, TX, USA)}
\begin{document}
\date{\vspace{-5ex}}
\maketitle
Comments: 34 pages, 11 figures.\\
\begin{abstract}
As the development of new classes of antibiotics slows, bacterial resistance to existing antibiotics is becoming an increasing problem. A potential solution is to develop treatment strategies with an alternative mode of action. We consider one such strategy: anti-adhesion therapy. Whereas antibiotics act directly upon bacteria, either killing them or inhibiting their growth, anti-adhesion therapy impedes the binding of bacteria to host cells. This prevents bacteria from deploying their arsenal of virulence mechanisms, while simultaneously rendering them more susceptible to natural and artificial clearance. In this paper, we consider a particular form of anti-adhesion therapy, involving biomimetic multivalent adhesion molecule (MAM) 7 coupled polystyrene microbeads, which competitively inhibit the binding of bacteria to host cells. We develop a mathematical model, formulated as a system of ordinary differential equations, to describe inhibitor treatment of a \emph{Pseudomonas aeruginosa} burn wound infection in the rat. Benchmarking our model against \emph{in vivo} data from an ongoing experimental programme, we use the model to explain bacteria population dynamics and to predict the efficacy of a range of treatment strategies, with the aim of improving treatment outcome. The model consists of two physical compartments: the epithelium and the exudate. It is found that, when effective in reducing the bacterial burden, inhibitor treatment operates both by preventing bacteria from binding to the epithelium and by reducing the flux of daughter cells from the epithelium into the exudate. Our model predicts that inhibitor treatment cannot eliminate the bacterial burden when used in isolation; however, when combined with regular or continuous debridement of the exudate, elimination is theoretically possible. Lastly, we present ways to improve therapeutic efficacy, as predicted by our mathematical model.
\end{abstract}
\section{Introduction}\label{Sec_Intro}
As we begin to lose the arms race against microbial infections, it is important that we develop new treatment strategies as a complement or alternative to antibiotics. In this paper, we use mathematical modelling to explain and predict the effects of a novel anti-adhesion therapy in the treatment of infected burn wounds, with the aim of improving treatment outcome.

Each year, millions of lives are saved through the use of antibiotics to combat bacterial infections. However, sustained use of any given antibiotic leads to the clinical emergence of drug-resistant strains. Since the discovery of penicillin, many new classes of antibiotics have been identified, allowing clinicians to switch between antibiotics if resistance emerges either within an individual patient or within a patient population \citep{Brannon_and_Hadjifrangiskou_2016}. Over time, strains have emerged which exhibit resistance to multiple classes of antibiotics (multi-drug resistance) and reports of bacterial infections which are resistant to all known antibiotics (pan-resistant) are becoming increasingly common. At present, a reported 700,000 individuals worldwide die each year due to antimicrobial resistance and this figure is predicted to rise to 10 million per year by 2050 unless steps are taken to combat this threat \citep{Rev_AMR_2016}. While resistant strains continue to evolve, our ability to develop new classes of antibiotics is diminishing, the rate of antibiotic discovery having slowed significantly since its `Golden Era' in the 1940s--1960s \citep{Brannon_and_Hadjifrangiskou_2016,Nathan_and_Cars_2014}. It is therefore vital that we develop alternative treatment strategies to replace or complement antibiotics \citep{Bush_et_al_2011,Teillant_et_al_2015}.

One potential way forward is through the use of anti-virulence treatments. Whereas antibiotics either kill bacteria (bactericidal) or inhibit their growth (bacteriostatic), anti-virulence treatments interfere with a pathogen's ability to cause damage and disease in the host \citep{Clatworthy_et_al_2007}. As such, they are likely to exert a smaller selective pressure upon a bacterial community, reducing the chances that resistance will develop \citep[though opinions vary over the extent to which they may be resistance-proof, see][]{Allen_et_al_2014,Vale_et_al_2014}. Anti-virulence treatments take a number of forms including those which target or inhibit toxin activity, adhesion, toxin secretion, virulence gene expression and inter-bacterial signalling \citep{Krachler_et_al_2012a,Krachler_and_Orth_2013,Rasko_and_Sperandio_2010}.

In this paper, we consider a form of anti-adhesion treatment consisting of polystyrene microbeads coupled to a protein known as multivalent adhesion molecule (MAM) 7 \citep[see also][and other papers from their group for alternative anti-adhesion treatments that operate by blocking pilus assembly or function]{Spaulding_et_al_2017}. MAM7 is anchored in the outer membrane of many Gram-negative bacteria, where it is responsible for initiating attachment of bacteria to host cells \citep{Krachler_et_al_2011_b,Krachler_and_Orth_2011}. When applied to an infection site, MAM7-coated beads (henceforth, inhibitors) act as a bacteriomimetic, competitively inhibiting the infectious agent from binding to host cells. This prevents bacteria from deploying those virulence mechanisms for which cell-to-cell contact is required and renders them more susceptible to natural or artificial physical clearance. Given that inhibitors must bind to host cells before bacteria in order to block them from binding, it is unclear whether their application may ever be expanded from prevention (prophylaxis) to the treatment of established infections (therapy).

Bacterial infection is a major cause of death in patients with burn wounds. `In patients with severe burns over more than 40\% of the total body surface area \ldots, 75\% of all deaths are currently related to sepsis from burn wound infection or other infection complications and/or inhalation injury' \citep{Church_et_al_2006}. Burn wounds are commonly infected by \emph{Pseudomonas aeruginosa} \citep{Azzopardi_et_al_2014,Church_et_al_2006,Oncul_et_al_2014,Pruitt_et_al_1998}, an opportunistic Gram-negative bacterium; the infection often being hospital-acquired (nosocomial) \citep{Church_et_al_2006,Weber_et_al_1997}.

Current treatment of such infections involves use of topical and systemic antibiotics, and regular debridement (mechanical wound cleaning). Debridement is either achieved through regular wound cleansing with a cloth, or through application of negative pressure devices (negative pressure wound therapy, NPWT) in which fluid is drawn out of the wound, either continuously or intermittently, using a pump, attached to a foam dressing covering the wound \citep{Moues_et_al_2004}. Some studies have shown NPWT to be effective in reducing the bacterial burden \citep{Mendonca_et_al_2006}; however, this result is not consistent across all studies \citep{Moues_et_al_2004,Vikatmaa_et_al_2008}.

In earlier work we have shown, using an experimental model for \emph{P.\ aeruginosa} burn wound infections in the rat, that treatment with inhibitors can significantly reduce the bacterial burden in the wound without impeding wound closure \citep[][see Section \ref{Sec_Exp_Set} for more details]{Huebinger_et_al_2016}. \emph{In vitro} studies have also demonstrated the efficacy of inhibitor treatment in reducing cytotoxicity \citep{Krachler_et_al_2012a,Krachler_et_al_2012b} and have shown that inhibitors do not interfere with host cell functions critical to wound healing \citep{Hawley_et_al_2013}.

A number of mathematical modelling studies have considered the use of anti-virulence treatments to combat bacterial infections. The majority of these studies focus upon anti-quorum sensing treatments \citep[see, for example,][]{Anguige_et_al_2004,Anguige_et_al_2005,Anguige_et_al_2006,Fagerlind_et_al_2005,Koerber_et_al_2002,Jabbari_et_al_2010,Jabbari_et_al_2012a,Jabbari_et_al_2012b}. An exception to this rule; the model by \citet{Ternent_et_al_2015} is of particular relevance to the present work. This ordinary differential equation (ODE) model considers a general anti-virulence treatment, which operates by enhancing innate immunity in bacterial clearance. The model predicts that, when used in isolation, anti-virulence treatment is unlikely to eliminate a bacterial infection. However, the model predicts that, when combined with antibiotics, anti-virulence treatments could eliminate bacteria, provided antibiotic and anti-virulence treatments are applied in staggered doses. Other modelling work has considered the bacterial invasion of burn wounds and the resultant tissue damage \citep{Hilhorst_et_al_2007a,Hilhorst_et_al_2007b,King_et_al_2003,Koerber_et_al_2002}, the influence of bacterial infection upon the healing of burn wounds \citep{Agyingi_et_al_2010} and the effects of ambient gas plasma treatment in this context \citep{Orazov_et_al_2012}. Each of these models is formulated as a system of partial differential equations in one or two spatial dimensions. Models have also been developed to describe microbial adhesion to surfaces, for example, \citet{Freter_et_al_1983} developed an ODE model for the competitive colonisation of the gut wall by host and invader strains of \emph{Escherichia coli}. Lastly, \citet{Gestel_and_Nowak_2016} have developed an individual-based model to describe the colonisation of a generic surface by phenotypically heterogeneous bacteria, in which bacteria may migrate between the surface and a liquid medium.

In this paper, we construct a mathematical model, formulated as a system of ODEs, to describe the population dynamics and treatment of a bacterial infection within a burn wound. Basing our mathematical model upon \citeauthor{Huebinger_et_al_2016}'s experiments, we use it to explain the empirical results and to predict the effects of various treatment regimes, involving inhibitor dosing and debridement, with the aim of improving efficacy. A particular strength of this study is that we consider multiple parameter sets, twelve in total, each of which provides a good fit to the experimental data. Classifying these sets into four qualitatively different cases, we consider the long-term effects of each treatment strategy, predicting the conditions under which treatment will eliminate the bacterial burden across all four cases.

The remainder of this paper is structured as follows: in Section \ref{Sec_Exp_Set} we describe \citeauthor{Huebinger_et_al_2016}'s experimental burn wound infection model, in Section \ref{Sec_Mod_Form} we construct a mathematical model to describe this experimental model and fit our mathematical model to the experimental data, in Section \ref{Sec_Results} we use our mathematical model to explain and predict the effects of various treatment strategies, lastly, in Section \ref{Sec_Disc} we discuss our findings and suggest directions for future research.
\section{Experimental set-up}\label{Sec_Exp_Set}
In this section we provide a simple description of the experimental set-up which forms the basis for our mathematical model. For a detailed description, see \citet{Huebinger_et_al_2016}.

\begin{figure}
\begin{center}
\includegraphics[scale=0.5]{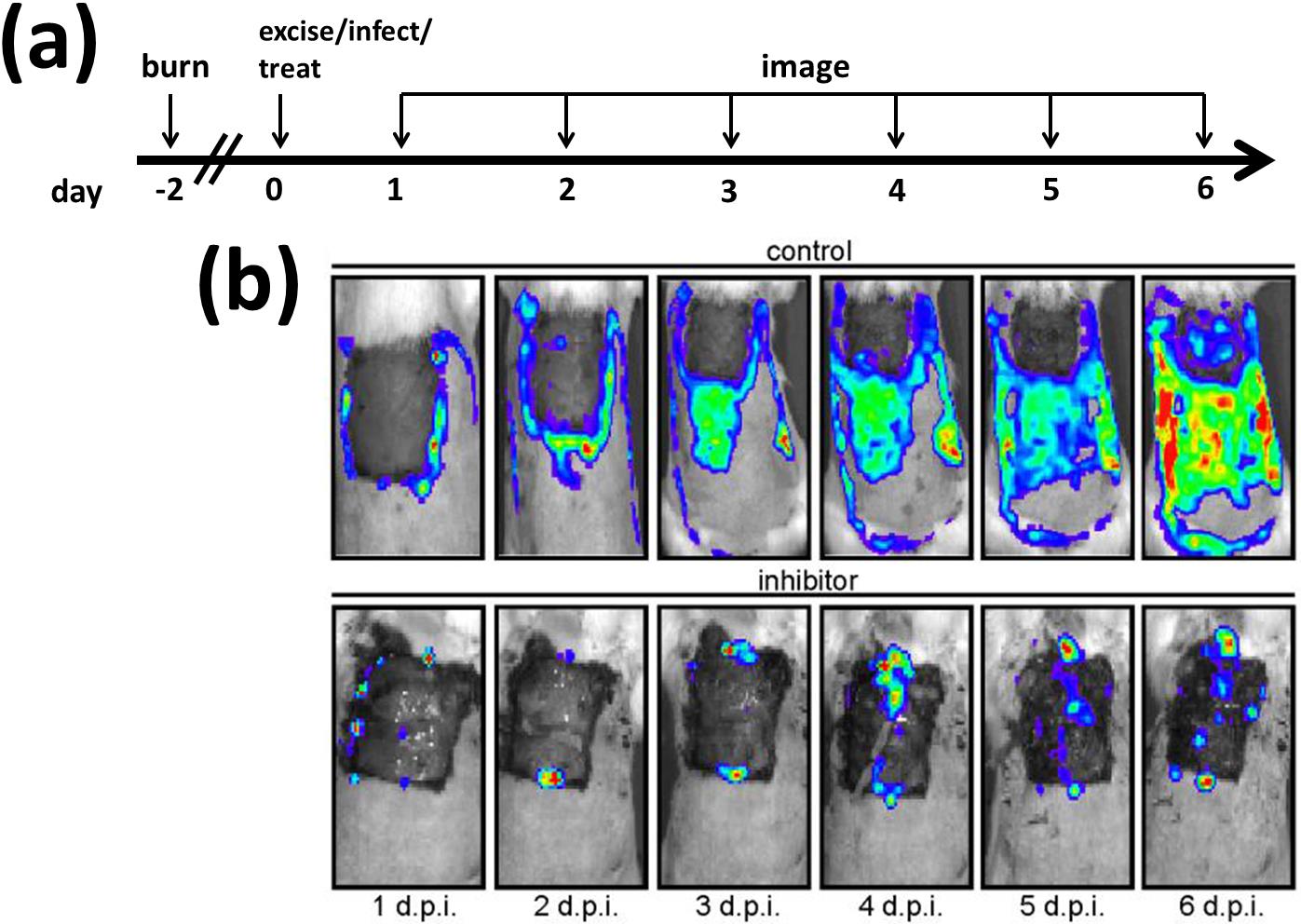}
\end{center}
\caption{Diagram summarising the experimental set-up. (a) Burns were administered to the backs of the rats on day $-2$. On day 0 a region of eschar (dead) tissue was excised and bioluminescent \emph{P.\ aeruginosa} were applied to the excision, followed by suspensions containing either inhibitor or control beads. Rats were imaged every 24 hours on days 1-6 post infection. (b) Exemplar images of burn wounds to which control beads (top) and inhibitor (bottom) were applied. In each image, the dark-grey area is the excision, the light-grey region is the non-excised burn wound, the white region is healthy, fur-covered tissue, and the coloured regions (colour online) denote the presence of bacteria, red corresponding to high density and blue to low density. The bacterial population size increases and spreads much more rapidly across the burn wound in the control scenario than when inhibitor is used. See Section \ref{Sec_Exp_Set} for more details. d\@.p\@.i\@.: days post infection. Figure reproduced, with modifications, from \citet{Huebinger_et_al_2016}.}
\label{Fig_Exp_Set}
\end{figure}

We consider a burn wound infection model in the Sprague-Dawley rat (see Fig.\ \ref{Fig_Exp_Set}). Rats were anaesthetised and the portion of each rat which was to be burned was shaved. Rats were then immersed in 100$^{\circ}$C water for 12s resulting in full-thickness cutaneous burns to 40\% of the body surface area, in a region spanning the back and upper sides of the body. We label the time at which the burn is administered as day $-2$. Rats were then resuscitated and given the appropriate pain control for the remainder of the experiment. On day 0, two days after the burn was administered, a section of eschar (dead) tissue, approximately 4 $\times$ 4 cm in area, was surgically excised. Next, 5 $\times$ 10$^6$ CFU (colony-forming units) of multidrug-resistant \emph{P.\ aeruginosa} were applied to the excised region, followed by suspensions containing either 3 $\times$ 10$^8$ inhibitor or control beads (without a MAM7 coating) in saline. Identical inhibitor and control treatments were repeated every 24 hours for days 1--5 post infection; however, since a scab (i\@.e.\ a layer of solidified exudate) forms over the excision by day 1, treatments administered on or after day 1 are unlikely to enter the (liquid) exudate. Rats were euthanized after the experiment, on day 6.

A bioluminescent, multidrug-resistant \emph{P.\ aeruginosa} isolate, Xen5, was chosen, such that the bacterial burden and their spatial distribution across the wound could be detected. An IVIS Spectrum \emph{in vivo} imaging system (Perkin Elmer) was used to record bacterial luminescence on days 1--6 post infection, from which the total flux (photons sec$^{-1}$) was calculated using MetaMorph software (Molecular Devices) to integrate over the pixels. The total number of bacteria in CFU was then calculated using the conversion factor 4 $\times$ 10$^8$ photons sec$^{-1}$ $\leftrightarrow$ 5 $\times$ 10$^6$ CFU, which was determined by measuring the luminescence of suspensions which contained an experimentally determined number of bacterial colony-forming units.

\begin{figure}
\begin{center}
\includegraphics[scale=0.6]{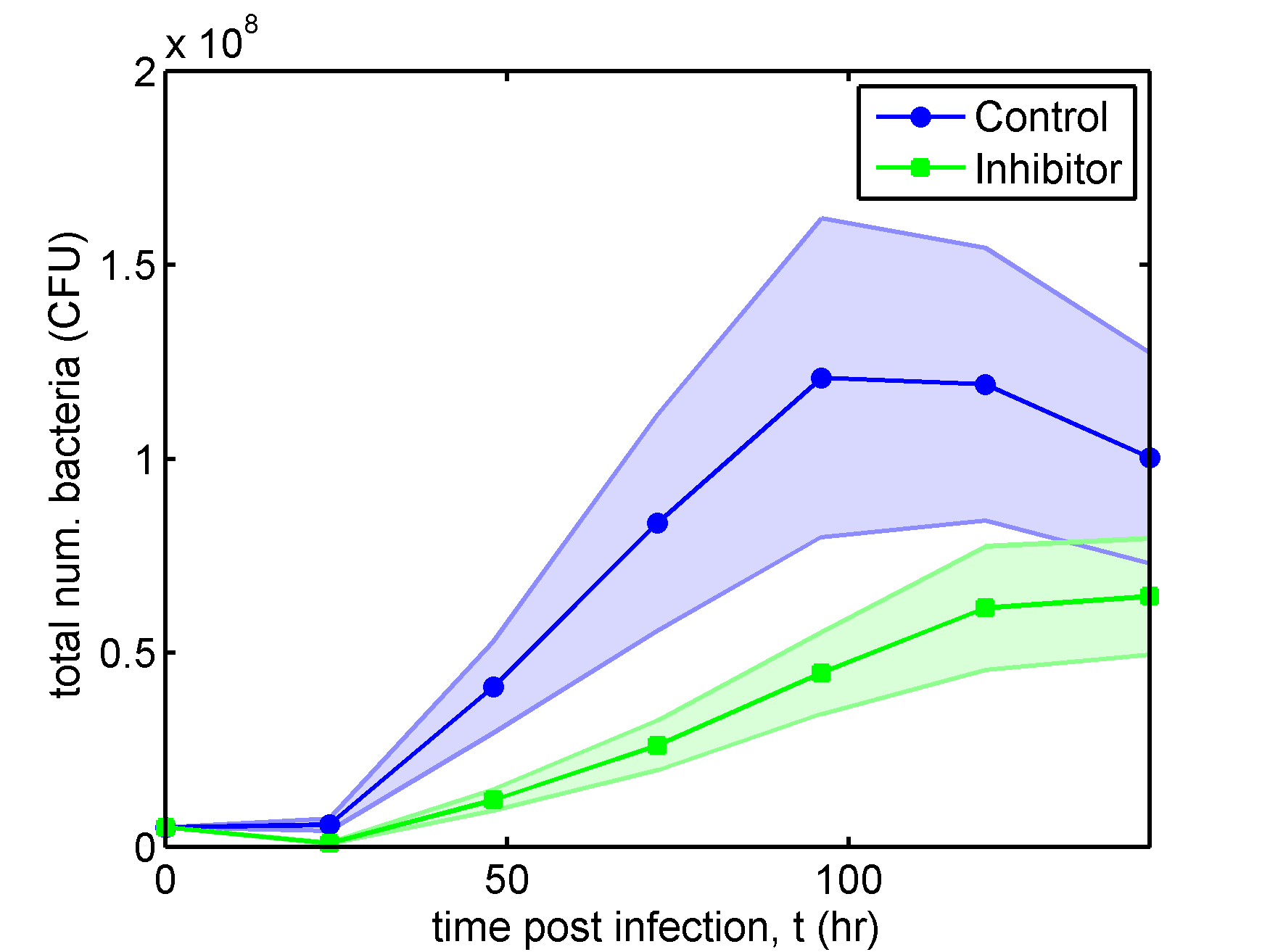}
\end{center}
\caption{Bacterial population size for control bead and inhibitor treated rats in the experimental model. The discs and squares show the mean bacterial population size in control bead ($n = 9$ animals) and inhibitor ($n = 13$ animals) treated rats respectively, while the shaded regions show the standard error of the mean. Treatment with the inhibitor reduces, but does not eliminate, the bacterial burden. CFU: colony-forming units.}
\label{Fig_Exp_Data}
\end{figure}

13 experiments were conducted using inhibitor and 11 using control beads. Two of the control bead experiments were discounted because the exposure setting used was too high to prevent the image from saturating. This may imply that the mean bacteria population size over time calculated for the control bead scenario slightly underestimates the true mean value. The experimental results are summarised in Fig.\ \ref{Fig_Exp_Data}.

These results raise two important questions:
\begin{enumerate}
\item \emph{Why is treatment with inhibitor effective in reducing the bacterial burden?} It is not clear why competitively inhibiting bacteria from binding to the epithelium should be as effective as it is in these experiments, since no physical clearance of free bacteria is employed.
\item \emph{How might the treatment be adapted to improve its efficacy?} For instance, this might be achieved by altering the adhesive properties of the inhibitor, the inhibitor dosing regimen or by combining inhibitors with other treatment strategies.
\end{enumerate}
In what follows, we formulate a mathematical model of the burn wound infection experiment described above to help us answer these questions.
\section{Model formulation}\label{Sec_Mod_Form}
\begin{figure}
\begin{center}
\includegraphics[scale=0.45]{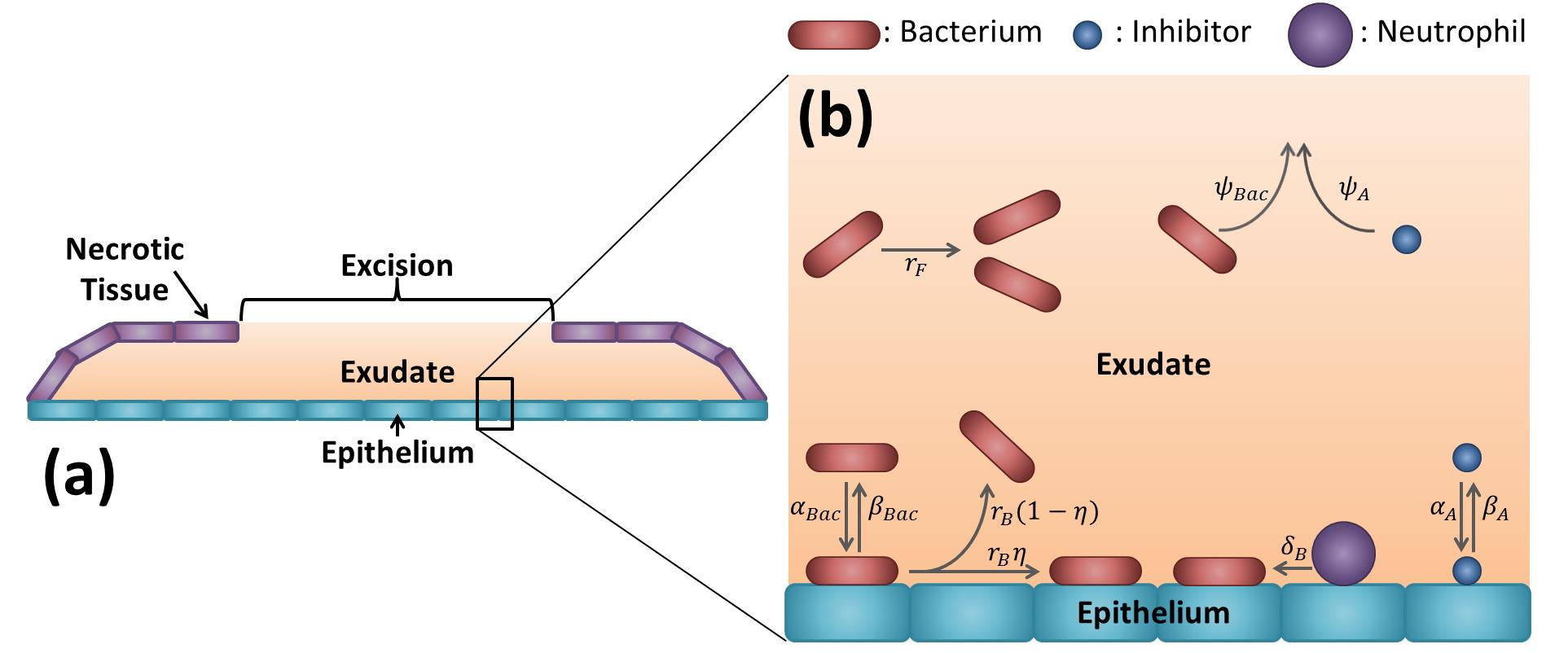}
\end{center}
\caption{Diagrams showing the wound geometry and model structure. (a) Wound geometry pictured in the transverse plane. The epithelium is covered by a liquid layer known as the exudate, which is itself covered by necrotic tissue, except in the region of the excision where the exudate is exposed to the air. (b) Diagram displaying the structure of the model given by Eqs.\ \eqref{Eqn_1}--\eqref{Eqn_7}. Bacteria and inhibitors exist in one of two states: free in the exudate or bound to the epithelium, and can transition between these states by binding to ($\alpha_{Bac}$ and $\alpha_A$) and unbinding from ($\beta_{Bac}$ and $\beta_A$) the surface. Both free and bound bacteria can divide ($r_F$ and $r_B$); daughters of free bacteria enter the exudate, whereas a proportion, $0 \leq \eta \leq 1$, of bound bacterial daughters remain bound to the surface, the remaining fraction, $1 - \eta$, entering the exudate. Bound bacteria may be phagocytosed by neutrophils ($\delta_B$), while free bacteria and inhibitors are cleared from the wound in the first 24 hrs after the excision is made and before a scab forms over the exposed exudate ($\psi_{Bac}$ and $\psi_A$).}
\label{Fig_Model_Diagram}
\end{figure}

The burn wound is assumed to consist of two physical compartments: the epithelium, and a fluid compartment exuded by the epithelium and (hence) known as the \emph{exudate}. The epithelium and the overlying exudate extend to the perimeter of the burn wound beneath the necrotic tissue, while the exudate is exposed to the air at the excision (see Fig.\ \ref{Fig_Model_Diagram}(a)). The area of the burn wound, $A$ (cm$^2$), remains essentially fixed during the experiment; however, the exudate height, $h$ (cm), and volume, $V$ (cm$^3$), are elevated for a short period following the application of bacteria and inhibitors to the excision at the beginning of day 0 (see Fig.\ \ref{Fig_Exp_Set}(a)). This excess fluid is lost rapidly via run-off (down the sides of the rat), evaporation and absorption (into the epithelium). Since the timescale over which the height and volume are elevated (on the order of minutes) is small compared to the timescale of the experiment (on the order of days), we neglect this variation and assume a fixed height and volume throughout the course of the experiment.

Both bacteria and inhibitors can exist in one of two states; either free in the exudate or bound to the epithelium. It is assumed that the system is well-mixed, allowing us to forgo an explicit spatial component and so to construct an ODE model for the evolution of the free bacteria density, $B_F(t)$ (cells cm$^{-3}$), bound bacteria density, $B_B(t)$ (cells cm$^{-2}$), free inhibitor concentration, $A_F(t)$ (inhibitors cm$^{-3}$), and bound inhibitor concentration, $A_B(t)$ (inhibitors cm$^{-2}$), over time, $t$ (hrs). It is assumed that the total binding site density on the epithelium, consisting of both free and occupied sites, is conserved, such that the free binding site density $E(t) = E_{init} - \phi_{Bac}B_B(t) - \phi_AA_B(t)$ (sites cm$^{-2}$), where $E_{init}$ (sites cm$^{-2}$) is the initial density of free binding sites, and $\phi_{Bac}$ (sites cell$^{-1}$) and $\phi_A$ (sites inhibitor$^{-1}$) are the number of binding sites occupied by a bacterium or an inhibitor respectively.

The model is summarised in Fig.\ \ref{Fig_Model_Diagram}(b) and described by the following governing equations
\begin{align}
\frac{\mathrm{d}B_F}{\mathrm{d}t} &= \underbrace{r_FB_F\left(1 - \frac{B_F}{K_F}\right)}_{\text{logistic growth}} + \underbrace{(1 - \eta(E))H(K_B-B_B)\frac{r_B}{h}B_B\left(1 - \frac{B_B}{K_B}\right)}_{\text{daughter cells freed from epithelium upon division}}
- \underbrace{\alpha_{Bac}AB_FE}_{\text{binding to epithelium}} + \underbrace{\frac{\beta_{Bac}}{h}B_B}_{\substack{\text{unbinding}\\\text{from epithelium}}}\nonumber\\
&\qquad - \underbrace{\psi_{Bac}(t)B_F}_{\text{natural clearance}}\text{,}\label{Eqn_1}\\
\frac{\mathrm{d}B_B}{\mathrm{d}t} &= \underbrace{(1 + (\eta(E) - 1)H(K_B-B_B))r_BB_B\left(1 - \frac{B_B}{K_B}\right)}_{\text{logistic growth (a proportion, $\eta$, remain attached)}}
+ \underbrace{\alpha_{Bac}VB_FE}_{\text{binding to epithelium}} - \underbrace{\beta_{Bac}B_B}_{\substack{\text{unbinding}\\\text{from epithelium}}} - \underbrace{\delta_BB_B}_{\text{phagocytosis}}\text{,}\label{Eqn_2}\\
\frac{\mathrm{d}A_F}{\mathrm{d}t} &= - \underbrace{\alpha_A A A_F E}_{\substack{\text{binding to}\\\text{epithelium}}} + \underbrace{\frac{\beta_A}{h}A_B}_{\substack{\text{unbinding}\\\text{from epithelium}}} - \underbrace{\psi_A(t)A_F}_{\substack{\text{natural}\\\text{clearance}}}\text{,}\label{Eqn_3}\\
\frac{\mathrm{d}A_B}{\mathrm{d}t} &= \underbrace{\alpha_A V A_F E}_{\substack{\text{binding to}\\\text{epithelium}}} - \underbrace{\beta_A A_B}_{\substack{\text{unbinding}\\\text{from epithelium}}}\text{,}\label{Eqn_4}
\end{align}
where parameter values are given in Tables \ref{Table_Param_1}, \ref{Table_Param_2} and S1.

Both free and bound bacteria are assumed to grow logistically with respective intrinsic growth rates $r_F$ (hr$^{-1}$) and $r_B$ (hr$^{-1}$), and carrying capacities $K_F$ (cells cm$^{-3}$) and $K_B$ (cells cm$^{-2}$). We interpret the carrying capacities to represent the maximum number of bacteria that can be sustained by available nutrients and the situation in which $B_F(t) = K_F$, or $B_B(t) = K_B$, to be one in which the rate of bacterial division is negligible \citep[see][]{Gefen_et_al_2014,Roostalu_et_al_2008}. We note that, in general, $K_B \neq E_{init}/\phi_{Bac}$, such that the number of bacteria that can be nourished on the epithelium is not equal to the number that can bind to the epithelium. For all of the parameter sets considered in this paper, $K_B < E_{init}/\phi_{Bac}$ (see Tables \ref{Table_Param_1}, \ref{Table_Param_2} and S1).

It is assumed that bacteria and inhibitors bind to and unbind from the epithelium in accordance with the law of mass action, with respective binding rates $\alpha_{Bac}$ (hr$^{-1}$ sites$^{-1}$) and $\alpha_A$ (hr$^{-1}$ sites$^{-1}$), and unbinding rates $\beta_{Bac}$ (hr$^{-1}$) and $\beta_A$ (hr$^{-1}$).

Examination of histological sections through the burn wound show that neutrophils are present within and at the surface of the epithelium, but not within the exudate \citep{Huebinger_et_al_2016}. Administration of a burn wound causes neutrophils to be fully activated such that no further neutrophils are recruited in response to the bacterial infection \citep{Agay_et_al_2008,Huebinger_et_al_2016} \citep[in contrast to][]{Ternent_et_al_2015}. Therefore, the immune response can be captured by the exponential decay of bound bacteria with rate $\delta_B$ (hr$^{-1}$), where $\delta_B$ accounts for the density of neutrophils. It is assumed that bead degradation, if it occurs, is sufficiently gradual that it can be neglected.

Several of the terms in Eqs.\ \eqref{Eqn_1}--\eqref{Eqn_4} contain $h$, $A$ or $V$ as a factor in order to ensure dimensional consistency. These constants could have been combined with other parameters, but we retain them in the interests of clarity.

A proportion of the daughter cells of bound bacteria, $0 \leq \eta(E(t)) \leq 1$ (dimensionless), remain bound to the surface, while the remaining fraction, $1 - \eta(E(t))$, enter the exudate. This proportion depends upon the density of free binding sites, $E(t)$, such that a larger fraction of the daughter cells remain bound when more binding sites are available. We capture this dependence using a Hill function with constant $\gamma$ (sites cm$^{-2}$) and Hill coefficient $n$ (dimensionless) as follows
\begin{equation}
\eta(E) = \frac{\eta_{max}E^n}{\gamma^n + E^n}\text{,}\label{Eqn_5a}
\end{equation}
where $\eta_{max}$ (dimensionless) is the maximum proportion of daughter cells which may remain bound to the surface. If the density of bound cells, $B_B(t)$, exceeds the bound carrying capacity, $K_B$, then the bound logistic growth term becomes a death term. In this case, the loss of bacteria is confined to the bound compartment and does not affect the free compartment. This is achieved through the use of a Heaviside step function, $H(K_B-B_B(t))$, in Eqs.\ \eqref{Eqn_1} and \eqref{Eqn_2}, where
\begin{align}
H(x) &\defeq \left\{
\begin{array}{rl}
0 & \text{if } x < 0\text{,}\\
1 & \text{if } x \geq 0\text{.}
\end{array}\right.
\label{Eqn_6}
\end{align}

The rates of clearance of bacteria and inhibitors, $\psi_{Bac}(t)$ (hr$^{-1}$) and $\psi_A(t)$ (hr$^{-1}$), vary with time, such that clearance occurs at a constant rate for the first 24 hours and then stops after this point due to the formation of a scab over the excision. Thus, clearance occurs at rates
\begin{equation}
\psi_{Bac}(t) = \tilde{\psi}_{Bac}H(24 - t)\text{\qquad and\qquad}\psi_A(t) = \tilde{\psi}_AH(24 - t)\text{,}\label{Eqn_5b}
\end{equation}
where $\tilde{\psi}_{Bac}$ (hr$^{-1}$) and $\tilde{\psi}_A$ (hr$^{-1}$) are the constant rates of clearance in the first 24 hours, and $H$ is a Heaviside step function, as defined in \eqref{Eqn_6}.

We choose the time $t = 0$ to correspond to the point at which bacteria and inhibitors are applied to the burn wound following the excision. Bacteria and inhibitors are present only in the free compartment initially, not having had the opportunity to bind to the epithelium, such that
\begin{equation}
B_F(0) = B_{F_{init}}\text{,\qquad} B_B(0) = 0\text{,\qquad} A_F(0) = A_{F_{init}}\text{,\qquad} A_B(0) = 0\text{,}\label{Eqn_7}
\end{equation}
where $B_{F_{init}}$ and $A_{F_{init}}$ are constants. See Tables \ref{Table_Param_1}, \ref{Table_Param_2} and S1 for parameter values.

The parameters in Tables \ref{Table_Param_1} and S1 were fitted to the experimental data (see Sections \ref{SubSec_Para_Fit} and S1 for details), while those in Table \ref{Table_Param_2} were either measured, calculated or estimated. The area of each burn wound was determined from images, such as those in Fig.\ \ref{Fig_Exp_Set}, using the MetaMorph software, while the height of the fluid layer has been measured to be 1 mm. As described in Section \ref{Sec_Exp_Set}, the initial density of bacteria and the initial concentration of inhibitor are known. The volume of the exudate is calculated as the product of the wound area and the height of the exudate. We know that there are about $1.5\times10^5$ epithelial cells per cm$^2$ and that approximately 17 inhibitors may bind per epithelial cell \citep{Krachler_et_al_2012a}. Taking the product of these two values gives us the initial density of free binding sites, $E_{init}$. Since we have defined the binding sites in terms of the space/receptors occupied by an inhibitor (rather than a bacterium), we have that $\phi_A = 1$ sites inhibitor$^{-1}$. Since inhibitors and bacteria are of a similar size, with a similar surface density of MAM7 molecules, we estimate that a bacterium occupies the same number of binding sites as an inhibitor. Simulations were found to be insensitive to the value of the Hill coefficient, $n$ ; therefore, we set it to unity for simplicity.

We leave our equations in dimensional form so as make them easier to interpret biologically and since non-dimensionalisation would not reduce the number of fitted parameters.
\begin{table}
\caption{Fitted parameter values for Eqs.\ \eqref{Eqn_1}--\eqref{Eqn_7}. The last three rows give the bacteria association constant, $\alpha_{Bac}/\beta_{Bac}$, the inhibitor association constant, $\alpha_A/\beta_A$, and the ratio of bacteria to inhibitor association constants. Values are given to an accuracy of 3 significant figures.}
\begin{center}
\begin{tabular}{l l p{1.2cm} p{1.2cm} p{1.2cm} p{1.2cm}}
\hline
\multirow{2}{*}{Parameter}	& \multirow{2}{*}{Description (Units)}										& \multicolumn{4}{c}{Value}\\ \cline{3-6}
							&																			& Case A					& Case B					& Case C					& Case D\\
\hline
$r_F$						& Intrinsic growth rate of free bacteria (hr$^{-1}$)						& 8.37$\times$10$^{-2}$		& 2.10$\times$10$^{-3}$		& 4.18$\times$10$^{-3}$		& 2.57$\times$10$^{-1}$\\
$r_B$						& Intrinsic growth rate of bound bacteria (hr$^{-1}$)						& 1.10$\times$10$^{-1}$		& 1.35$\times$10$^{-1}$		& 1.50$\times$10$^{-1}$		& 5.55\\
$K_F$						& Carrying capacity of free bacteria (cells cm$^{-3}$)						& 1.17$\times$10$^{7}$		& 1.15$\times$10$^{6}$		& 3.12$\times$10$^{6}$		& 1.85$\times$10$^{6}$\\
$K_B$ 						& Carrying capacity of bound bacteria (cells cm$^{-2}$)						& 9.96$\times$10$^{5}$		& 1.65$\times$10$^{6}$		& 1.48$\times$10$^{6}$		& 1.43$\times$10$^{6}$\\
$\alpha_{Bac}$				& Binding rate of bacteria to epithelium (hr$^{-1}$ sites$^{-1}$)			& 1.34$\times$10$^{-9}$		& 3.09$\times$10$^{-10}$	& 2.32$\times$10$^{-10}$	& 3.34$\times$10$^{-11}$\\
$\beta_{Bac}$				& Unbinding rate of bacteria from epithelium (hr$^{-1}$)					& 1.97$\times$10$^{-1}$		& 4.11$\times$10$^{-9}$		& 8.23$\times$10$^{-9}$		& 5.79$\times$10$^{-6}$\\
$\delta_B$					& Rate of phagocytosis of bacteria by neutrophils (hr$^{-1}$)				& 1.06$\times$10$^{-3}$		& 2.31$\times$10$^{-4}$		& 3.97$\times$10$^{-4}$		& 3.02$\times$10$^{-5}$\\
$\eta_{max}$ 				& Maximum proportion of daughters of bound cells 							& 2.95$\times$10$^{-2}$		& 1.31$\times$10$^{-7}$		& 3.06$\times$10$^{-7}$		& 1.52$\times$10$^{-2}$\\
							& that can be released into the exudate (dimensionless)						& 							& 							&							&\\
$\gamma$ 					& Concentration of binding sites at which 									& 3.12$\times$10$^{4}$		& 1.83$\times$10$^{5}$		& 3.66$\times$10$^{4}$		& 1.65$\times$10$^{6}$\\
							& $\eta = \eta_{max}/2$	(sites cm$^{-2}$)									& 							& 							& 							&\\
$\tilde{\psi}_{Bac}$ 		& Natural clearance rate of bacteria (hr$^{-1}$)							& 1.42$\times$10$^{-1}$		& 2.13$\times$10$^{-6}$		& 7.68$\times$10$^{-6}$		& 5.01$\times$10$^{-1}$\\
$\alpha_A$ 					& Binding rate of inhibitors to epithelium (hr$^{-1}$ sites$^{-1}$)				& 1.46$\times$10$^{-6}$		& 1.56$\times$10$^{-10}$	& 2.32$\times$10$^{-10}$	& 5.51$\times$10$^{-9}$\\
$\beta_A$ 					& Unbinding rate of inhibitors from epithelium (hr$^{-1}$)						& 6.35$\times$10$^{-8}$		& 4.11$\times$10$^{-9}$		& 3.85$\times$10$^{-3}$		& 4.43$\times$10$^{-1}$\\
$\tilde{\psi}_A$ 			& Natural clearance rate of inhibitors (hr$^{-1}$)								& 4.39$\times$10$^{-8}$		& 2.13$\times$10$^{-6}$		& 3.85$\times$10$^{-3}$		& 1.75$\times$10$^{-5}$\\
$\alpha_{Bac}/\beta_{Bac}$ 	& Bacterial association constant (sites$^{-1}$)								& 6.82$\times$10$^{-9}$		& 7.53$\times$10$^{-2}$		& 2.82$\times$10$^{-2}$		& 5.77$\times$10$^{-6}$\\
$\alpha_A/\beta_A$ 			& Inhibitor association constant (sites$^{-1}$)									& 2.29$\times$10$^{1}$		& 3.79$\times$10$^{-2}$		& 6.03$\times$10$^{-8}$		& 1.24$\times$10$^{-8}$\\
$\frac{(\alpha_{Bac}/\beta_{Bac})}{(\alpha_A/\beta_A)}$ & Ratio of association constants (dimensionless)& 2.97$\times$10$^{-10}$	& 1.98						& 4.68$\times$10$^{5}$		& 4.65$\times$10$^{2}$\\
\hline
\end{tabular}
\end{center}
\label{Table_Param_1}
\end{table}
\begin{table}
\caption{Measured, calculated and estimated parameter values for Eqs.\ \eqref{Eqn_1}--\eqref{Eqn_7}. Measured values are those which have been measured directly, calculated values are those which have been calculated using values which were measured directly and estimated values are those which could not be measured or calculated. The values in brackets are used only when fitting the inhibitor treatment data to the model (Figs.\ \ref{Fig_Model_and_Data} and S2).}\label{Table_Param_2}
\begin{center}
\begin{tabular}{llll}
\hline
Parameter 				& Description (Units)													& Value										& Source\\
\hline
$\phi_{Bac}$			& Number of binding sites occupied by a bacterium (sites cell$^{-1}$)	& 1 										& Estimated\\
$\phi_A$				& Number of binding sites occupied by an inhibitor (sites inhibitor$^{-1}$)		& 1											& Calculated\\
$V$						& Volume of the exudate	(cm$^3$)										& 4.9 (4.6)									& Calculated\\
$A$						& Area of the burn wound (cm$^2$)	 									& 49 (46)									& Measured\\
$h$						& Height of the exudate	(cm)											& 0.1										& Measured\\
$n$						& Hill coefficient (dimensionless)										& 1											& Estimated\\
$B_{F_{init}}$			& Initial density of free bacteria (cells cm$^{-3}$)					& 1.02$\times$10$^6$ (1.09$\times$10$^6$)	& Measured\\
$A_{F_{init}}$			& Initial concentration of free inhibitors (inhibitors cm$^{-3}$)					& 0 or 6.12$\times$10$^7$(6.52$\times$10$^7$)& Measured\\
$E_{init}$				& Initial density of binding sites (sites cm$^{-2}$)					& 2.57$\times$10$^6$						& Calculated\\
\hline
\end{tabular}
\end{center}
\end{table}
\subsection{Treatment scenarios}\label{SubSec_Treat_Strat}
In addition to the untreated/control ($A_{F_{init}}=0$) and single inhibitor dose ($A_{F_{init}}>0$) scenarios based upon \citeauthor{Huebinger_et_al_2016}'s experiments (see Section \ref{Sec_Exp_Set}), we consider a further six theoretical scenarios, five of which include either regular or continuous debridement. Since inhibitors operate by blocking bacteria from binding to the wound epithelium, it is intuitive that this may result in the majority of bacteria occupying the free compartment. Thus, any treatment, such as debridement, which removes the exudate, could clear the free compartment of bacteria --- and with them, inhibitors --- significantly reducing the total population size of bacteria when combined with an inhibitor treatment.

Regular debridement consists of a series of discrete instantaneous debridement events, while continuous debridement consists of a sustained, high level of clearance ($\psi_{Bac} = \psi_A = 1000$ hr$^{-1}$) and may be thought of as the limiting case of regular debridement in which the time between debridement events tends to zero. While it may not be possible to maintain such a high rate of clearance in practice, this allows us to determine the theoretical best-case-scenario were such a treatment applied. Following a discrete debridement event, it is assumed that the fluid compartment is restored on the timescale of a few minutes, such that the volume fluctuation can be neglected.

The eight treatment scenarios may be summarised as follows:
\begin{enumerate}
\item \emph{No treatment}: no inhibitors or debridement are applied;
\item \emph{Single inhibitor dose}: inhibitors are applied at time $t = 0$ hr;
\item \emph{Regular inhibitor doses}: inhibitors are applied at time $t = 0$ hr, then again at $t = 48$ hr and reapplied every 24 hr;
\item \emph{Regular debridement}: all free bacteria and inhibitors are removed at $t = 48$ hr and the procedure repeated every 24 hr;
\item \emph{Single inhibitor dose with regular debridement}: inhibitors are applied at time $t = 0$ hr, after which all free bacteria and inhibitors are removed at $t = 48$ hr and the debridement event repeated every 24 hr;
\item \emph{Regular inhibitor doses with regular debridement}: inhibitors are applied at time $t = 0$ hr, and free bacteria and inhibitors are removed at $t = 48$ hr directly after which inhibitors are reapplied, the debridement-inhibitor treatment is then repeated every 24 hrs;
\item \emph{Continuous debridement}: all free bacteria and inhibitors are removed at $t = 48$ hr, after which clearance of bacteria and inhibitors is maintained at a high level;
\item \emph{Single inhibitor dose with continuous debridement}: inhibitors are applied at time $t = 0$ hr, and all free bacteria and inhibitors are removed at $t = 48$ hr, after which clearance of bacteria and inhibitors is maintained at a high level.
\end{enumerate}
For each of the treatment strategies, Eqs.\ \eqref{Eqn_1}--\eqref{Eqn_7} were solved using the Matlab routine \textsf{ode15s}, a variable-step, variable-order solver based upon numerical differentiation formulas.

The untreated and single inhibitor dose scenarios are those considered in the experiments and are described above and in Section \ref{Sec_Exp_Set}. The key difference between the numerical simulations in Section \ref{SubSec_NumSol} and the experiments is that the simulations extend beyond the time frame of the experiments.

In the regular inhibitor dose scenario (and the regular inhibitor dose with regular debridement scenario), the repeat doses of inhibitors are identical to the initial dose, $A_{F_{init}}$. The second dose is not applied until 48 hrs for consistency with the treatments involving debridement (see below). As noted in Section \ref{Sec_Exp_Set}, a scab forms over the wound after the first 24 hrs, ending clearance and preventing further inhibitor doses from reaching the exudate. Thus, in practice, inhibitor doses could not be repeated without removing the scab and incurring further clearance at levels similar to those in the first 24 hrs. However, since we are interested in the theoretical effect of repeated inhibitor doses independent of clearance, we neglect further clearance effects in this case. Were we to include additional clearance upon re-treatment with inhibitor, treatment efficacy would be improved.

In the scenarios involving regular debridement, clearance is re-established for the first 24 hours after each debridement event, with rates given in Tables \ref{Table_Param_1} and S1, to account for leakage due to the loss of the scab upon debridement. The first debridement event is chosen to occur at $t = 48$ hr, rather than some earlier time, so as to give the inhibitors time to bind to the epithelium.
\subsection{Sensitivity analyses}\label{SubSec_SA}
We present two sets of sensitivity analyses. The first set (presented in Sections \ref{SubSubSec_Case_A}--\ref{SubSubSec_Case_D}) shows the effect of a tenfold increase or decrease in each of the 13 fitted parameters, $r_F$, $r_B$, $K_F$, $K_B$, $\alpha_{Bac}$, $\beta_{Bac}$, $\delta_B$, $\eta_{max}$, $\gamma$, $\tilde{\psi}_{Bac}$, $\alpha_A$, $\beta_A$ and $\tilde{\psi}_A$, on the total number of bacteria either at steady-state (treatment scenarios 1 and 2) or at $t = 90$ days $= 2160$ hrs (treatment scenarios 3--8; Figs.\ S15--S22). We truncate the simulations for the latter treatment scenarios at 90 days since simulating treatments with regular inhibitor doses or regular debridement is computationally expensive and those involving continuous debridement are close to steady-state by this time. We note that the total population size of bacteria oscillates in those treatments that involve regular debridement, undergoing  a sharp drop upon each debridement event. We plot the value of $B_T(t)$ at the peak of the oscillation at $t = 90$ days (directly prior to debridement), since we consider that it is by the maximum number of bacteria that the efficacy of a treatment should be judged.

In the second set of sensitivity analyses (presented in Section \ref{SubSubSec_2DSA}), we consider the effect of varying the binding and unbinding rates of inhibitors, $\alpha_A$ and $\beta_A$, in the space $\{10^{-12},10^{-11},\ldots ,1\}\times\{10^{-12},10^{-11},\ldots ,1\}$ upon the total number of bacteria at 4 weeks ($= 672$ hr) post infection in the 5 scenarios that involve inhibitor treatment (Figs.\ \ref{Fig_2D_SA}, S23, S25, S27, S29 and S31). We also consider the effect of increasing all inhibitor doses by 10 fold from $6.12\times10^7$ inhibitors cm$^{-3}$ (the standard value) to $6.12\times10^8$ inhibitors cm$^{-3}$ (Figs.\ S24, S26, S28, S30 and S32).
\subsection{Parameter fitting}\label{SubSec_Para_Fit}
A combination of Markov Chain Monte Carlo (MCMC) and  frequentist methods were used to fit the model given by Eqs.\ \eqref{Eqn_1}--\eqref{Eqn_7} to the mean of the experimental data in the untreated and single inhibitor dose scenarios. There is insufficient data presently available to establish a unique model fit; however, we are able to identify a number of good fits (twelve parameters sets are presented here) and to classify these into four general cases --- A, B , C and D --- based upon their qualitative behaviour (see Section \ref{Sec_Results}). By considering a range of valid parameter sets, rather than a single best fit, we are able to gain a more comprehensive understanding of the model behaviour. The fitting procedures used differ between parameter sets and are summarised in Table S2. See Section S1 of the Supplementary Materials for more details.

\begin{figure}
\begin{center}
\includegraphics[scale=0.59]{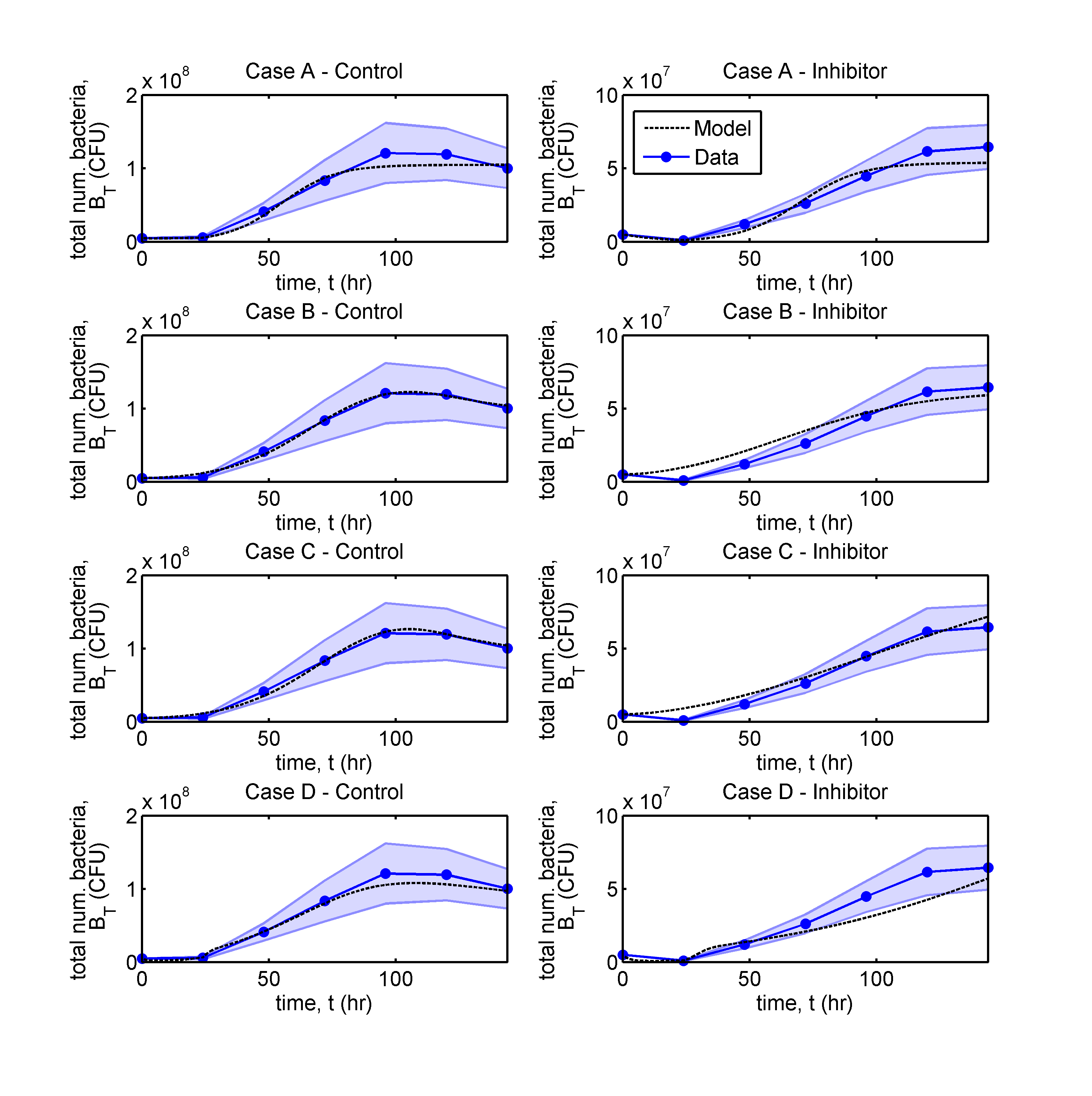}
\end{center}
\caption{Comparison of model predictions with experimental data. The discs mark the experimental mean, while the shaded region shows the standard error of the mean. Simulation results, denoted by the dashed line, show the total number of bacteria, $B_T(t)$ ($= VB_F(t) + AB_B(t)$). Note that the range of the $y$-axis for the untreated (control) scenario (left-hand column) is twice that for the single inhibitor dose scenario (right-hand column). There is good agreement between the model and the data in all cases and the overall goodness of fit (mean squared error) is similar for each parameter set.  Eqs.\ \eqref{Eqn_1}--\eqref{Eqn_7} were solved using \textsf{ode15s} and fitting was performed using a combination of MCMC and frequentist methods (see Sections \ref{SubSec_Para_Fit} and S1). See Tables \ref{Table_Param_1} and \ref{Table_Param_2} for parameter values. CFU: colony-forming units.}
\label{Fig_Model_and_Data}
\end{figure}

Model fits are compared against the experimental data in Figs. \ref{Fig_Model_and_Data}, S1 and S2, where parameter sets 2 (Case A), 3 (Case B), 8 (Case C) and 12 (Case D) are presented in Fig. \ref{Fig_Model_and_Data}. Since the experimental data does not distinguish between free and bound bacteria, we compare it against the simulated total number of bacteria, $B_T(t) = VB_F(t) + AB_B(t)$. The model achieves a good fit to the experimental mean in all cases, remaining mostly within the shaded region denoting the standard error of the mean.
\section{Results}\label{Sec_Results}
In what follows, we examine the behaviour of Eqs.\ \eqref{Eqn_1}--\eqref{Eqn_7} in each of the four cases, A--D. For clarity, the results from four parameter sets, one from each case, are presented in the main text: Set 2 from Case A, Set 3 from Case B, Set 8 from Case C and Set 12 from Case D. For the complete set of results, see Section S2 of the Supplementary Materials. We begin with a steady-state analysis of the system with and without a single inhibitor dose in order to determine the number of steady-states and their stability properties. We then consider the behaviour of the time-dependent problem, simulating the bacterial population dynamics well beyond the time frame of the experiments. While simulations provide a good fit to the experimental data for all parameter sets, the model predictions diverge for later times. We investigate why a single inhibitor dose is or is not effective in each case in the long-term and explore other potential treatment strategies with the aim of improving efficacy. We consider a treatment to be effective if it reduces the bacterial burden and to be fully effective if the bacterial burden is eliminated (such that $B_T(t) < 1$).
\subsection{Steady-state analysis}\label{SubSec_StStAnal}
Steady-state analyses of Eqs.\ \eqref{Eqn_1}--\eqref{Eqn_6}, in the absence of clearance, with and without a single dose of inhibitors were performed using Maple. Clearance was neglected since leakage of fluid from the wound only occurs in the first 24 hours. It was found that the system has two physically realistic steady-states in both the untreated and single inhibitor dose scenarios for all 12 parameter sets. In the untreated scenario, the first steady-state, $(B_{F_1}^*,B_{B_1}^*) = (0,0)$, corresponds to the complete absence of bacteria and can be classified as a saddle-node in all cases, with an unstable manifold directed into the positive quadrant. For the second steady-state, $(B_{F_2}^*,B_{B_2}^*)$, we have that $B_{F_2}^* > 0$ and $B_{B_2}^* > 0$ in all cases. Depending upon the parameter set, this steady-state takes the form of either a stable improper node or a stable spiral (see Table S1).

In the treated scenario, since we are neglecting clearance, the total number of inhibitors ($A_T = VA_F(t) + AA_B(t) = VA_{F_{init}}$) is conserved. Therefore, we may substitute for $A_B(t)$ as $A_B(t) =  h(A_{F_{init}} -A_F(t))$ into Eqs.\ \eqref{Eqn_1}--\eqref{Eqn_4}, reducing the problem to the following three-dimensional system:
\begin{align}
\frac{\mathrm{d}B_F}{\mathrm{d}t} &= r_FB_F\left(1 - \frac{B_F}{K_F}\right) + (1 - \eta(E))H(K_B-B_B)\frac{r_B}{h}B_B\left(1 - \frac{B_B}{K_B}\right) - \alpha_{Bac}AB_FE + \frac{\beta_{Bac}}{h}B_B\text{,}\label{Eqn_8}\\
\frac{\mathrm{d}B_B}{\mathrm{d}t} &= (1 + (\eta(E) - 1)H(K_B-B_B))r_BB_B\left(1 - \frac{B_B}{K_B}\right)
+ \alpha_{Bac}VB_FE - \beta_{Bac}B_B - \delta_BB_B\text{,}\label{Eqn_9}\\
\frac{\mathrm{d}A_F}{\mathrm{d}t} &= - \alpha_A A A_F E + \beta_A(A_{F_{init}} -A_F)\text{.}\label{Eqn_10}
\end{align}

\begin{figure}
\begin{center}
\includegraphics[scale=0.525]{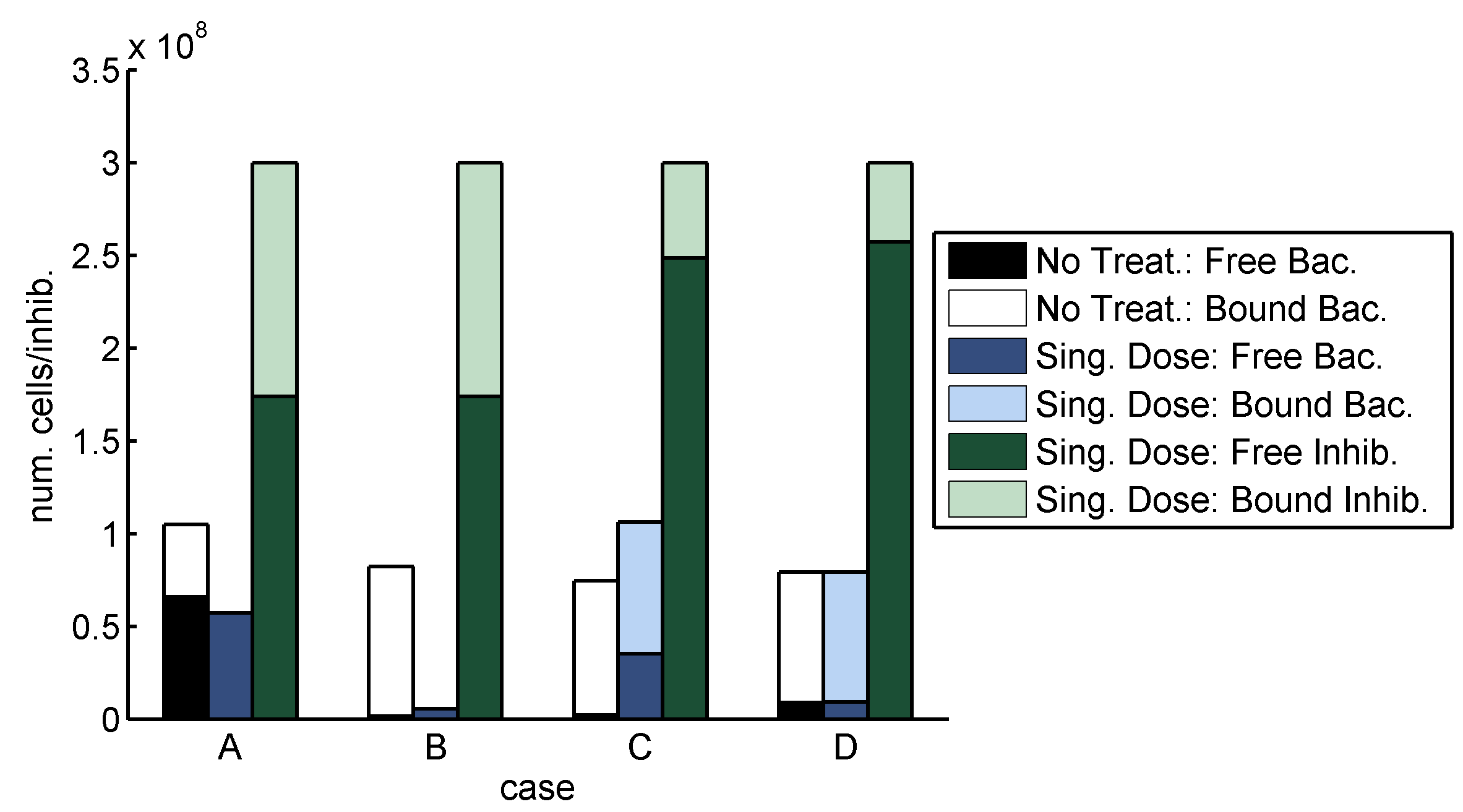}
\end{center}
\caption{Steady-state solutions to Eqs.\ \eqref{Eqn_1}--\eqref{Eqn_6} with and without a single dose of inhibitor. Three stacked bars are plotted for each case: the first bar shows the number of free and bound bacteria, $\hat{B}_{F_2}^* = VB_{F_2}^*$ and $\hat{B}_{B_2}^* = AB_{B_2}^*$, at steady-state in the untreated scenario; the second gives the number of free and bound bacteria at steady-state in the single inhibitor dose scenario, and the third shows the number of free and bound inhibitors, $\hat{A}_{F_2}^* = VA_{F_2}^*$ and $\hat{A}_{B_2}^* = AA_{B_2}^*$, at steady-state in the single inhibitor dose scenario. The combined height of each stacked bar gives the total number of bacteria or inhibitors, $B_{T_2}^* = \hat{B}_{F_2}^* + \hat{B}_{B_2}^*$ and $A_{F_{init}} = \hat{A}_{F_2}^* + \hat{A}_{B_2}^*$. Treatment results in a decrease in $B_{T_2}^*$ in Cases A and B, an increase in $B_{T_2}^*$ in Case C and has little effect on $B_{T_2}^*$ in Case D. The ratio of free to bound inhibitors is similar in Cases A and B, and varies in Cases C and D. Steady-state solutions were obtained by solving Eqs.\ \eqref{Eqn_1}--\eqref{Eqn_7} using \textsf{ode15s}, allowing the system to evolve until it reached steady-state. The problem was solved in the absence of clearance, such that $\tilde{\psi}_{Bac} = 0$ and $\tilde{\psi}_A = 0$. See Tables \ref{Table_Param_1} and \ref{Table_Param_2} for the remaining parameter values.}
\label{Fig_St_St}
\end{figure}

This system has two steady-states; the first, $(B_{F_1}^*,B_{B_1}^*,A_{F_1}^*) = (0,0,A_{F_1}^*)$, where $A_{F_1}^* > 0$, corresponds to the complete absence of bacteria and can be classified as a saddle node (with either one or two unstable manifolds) in all cases. For the second steady-state, $(B_{F_2}^*,B_{B_2}^*,A_{F_2}^*)$, we have that $B_{F_2}^* > 0$, $B_{B_2}^* > 0$ and $A_{F_2}^* > 0$ in all cases, though $B_{B_2}^*$ is tiny by comparison with $B_{F_2}^*$ and $A_{F_2}^*$ for Cases A and B (see Figs.\ \ref{Fig_St_St} and S3, where $B_{B_2}^*$ is too small to be visible in Cases A and B). This steady-state takes the form of a stable improper node (three non-identical real negative eigenvalues) for all parameter sets, except Set 8, where it is a stable spiral (one real negative eigenvalue and a pair of complex conjugate eigenvalues with negative real part).

By characterising the stability of the system in this way, we can be sure that we are not overlooking any potential stable steady-state solutions in the time-dependent simulations below. The discovery that the stable steady-state may take the form of a node or a spiral may also be of value in identifying the most accurate parameter set, if it were possible to take highly resolved experimental measurements, so as to detect, or rule out, oscillations in the bacterial population size.

The steady-state solutions in the untreated and single inhibitor dose scenarios are summarised in Figs.\ \ref{Fig_St_St} and S3. Treatment results in a decrease in the total number of bacteria in Cases A and B, as might be expected from the experimental results (see Fig.\ \ref{Fig_Exp_Data}); however, it results in an increase in Case C and has little effect on the total population size in Case D. These latter two results are unexpected, suggesting that in some situations treatment with inhibitor could be detrimental in the long-term. We discuss this further in Section \ref{SubSec_NumSol} below. The ratio of free inhibitors to bound inhibitors is consistent across all parameter sets in Cases A and B, but varies across parameter sets in Cases C and D, with a lower proportion of bound inhibitors in Cases C and D.
\subsection{Numerical solutions}\label{SubSec_NumSol}
Having explored the behaviour of the system at steady-state, we consider the full time-dependent problem (Eqs.\ \eqref{Eqn_1}--\eqref{Eqn_7}). Taking Cases A--D in turn, we note the distinguishing features of each case, explain the solution behaviour in the untreated and single inhibitor dose scenarios, and investigate other treatment strategies. 

\begin{figure}
\begin{center}
\includegraphics[scale=0.69]{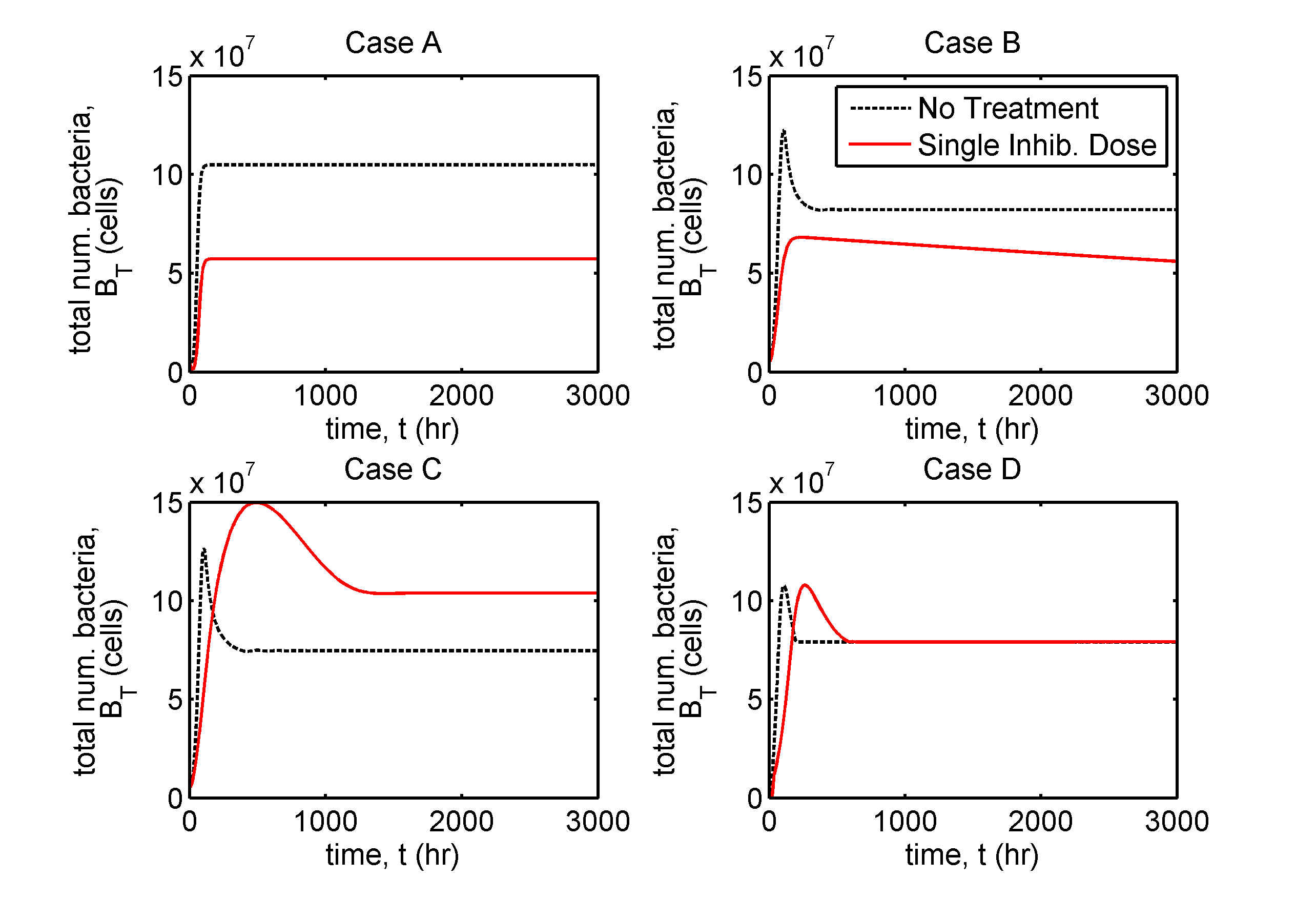}
\end{center}
\caption{Comparison of bacterial population dynamics in the untreated and single inhibitor dose scenarios. The total number of bacteria, $B_T(t)$ ($= VB_F(t) + AB_B(t)$), is plotted in each case. Simulations extend beyond the time span of the experiments, to 3000 hours $=$ 125 days. Case A: the bacterial population size is an essentially monotone increasing function of time in both untreated and treated scenarios, reaching an early steady-state at which the bacterial population size with treatment is roughly half that without; Case B: the bacterial population size with treatment remains below that without treatment, decreasing gradually after reaching an early maximum; Case C: treatment results in a sustained and significant increase in the bacterial population size, Case D: treatment causes the bacterial population size to temporarily exceed that without treatment, settling to a steady-state close to that of the untreated scenario. Eqs.\ \eqref{Eqn_1}--\eqref{Eqn_7} were solved using \textsf{ode15s}. See Tables \ref{Table_Param_1} and \ref{Table_Param_2} for parameter values.}
\label{Fig_NT_and_Treat}
\end{figure}

\begin{figure}
\begin{center}
\includegraphics[scale=0.59]{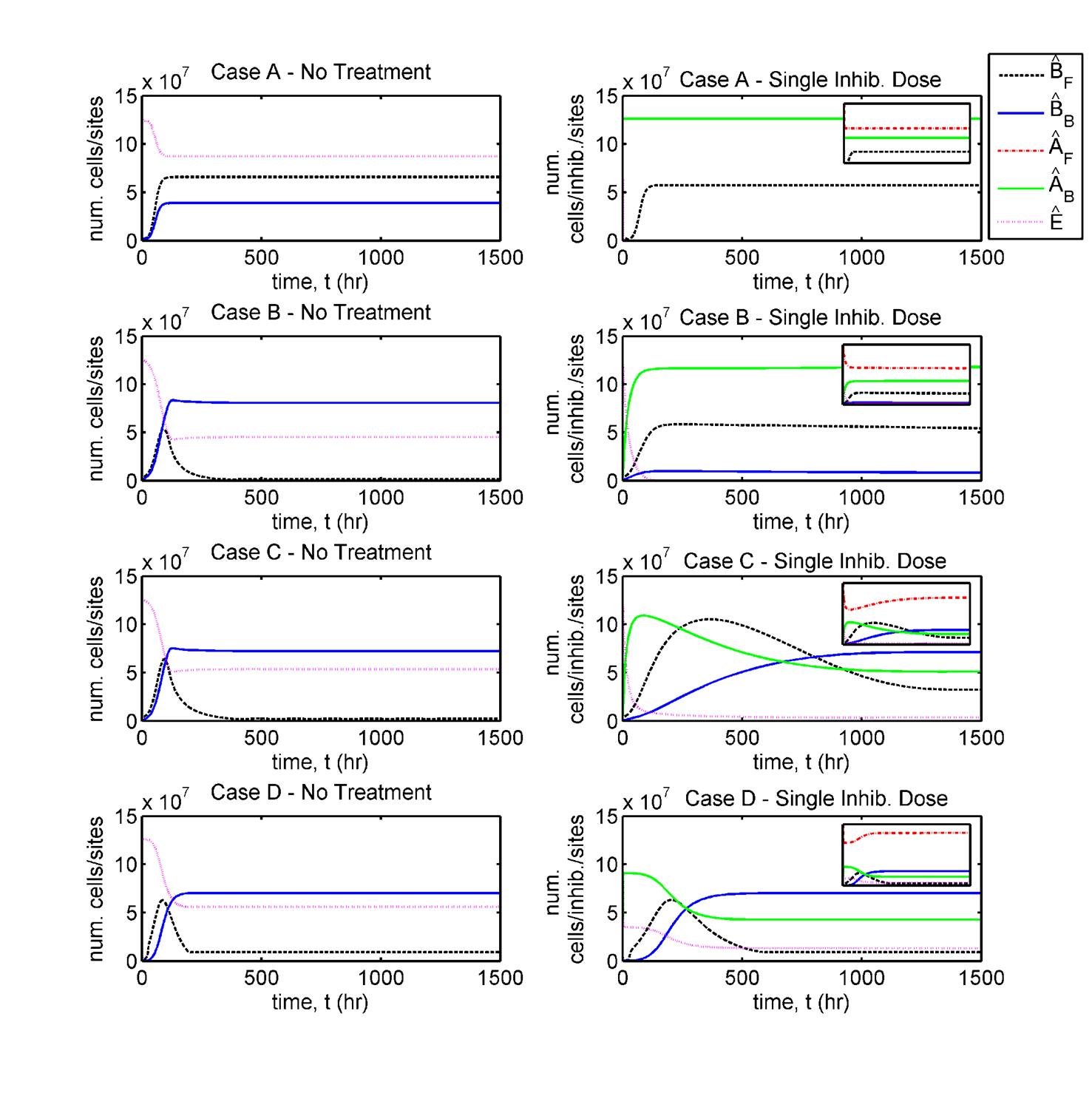}
\end{center}
\caption{Comparison of dependent variables with and without a single dose of inhibitor. The total number of cells ($\hat{B}_F(t) = VB_F(t)$, $\hat{B}_B(t) = AB_B(t)$), inhibitors ($\hat{A}_F(t) = VA_F(t)$, $\hat{A}_B(t) = AA_B(t)$) and binding sites ($\hat{E}(t) = AE(t)$) are plotted in each case. Simulations extend beyond the time span of the experiments, to 1500 hours $=$ 62.5 days.  The insets in the right-hand column show the same plots as in the larger panels, but with the $y$-axis spanning $[0,3\times 10^8]$. Cases A and B: the number of bound bacteria are reduced as a result of treatment; Case C: treatment results in an increase in the number of free bacteria; Case D: treatment increases the time taken to reach a roughly equivalent steady-state to that without treatment. Eqs.\ \eqref{Eqn_1}--\eqref{Eqn_7} were solved using \textsf{ode15s}. See Tables \ref{Table_Param_1} and \ref{Table_Param_2} for parameter values.}
\label{Fig_Dep_Var}
\end{figure}

\begin{figure}
\begin{center}
\includegraphics[scale=0.59]{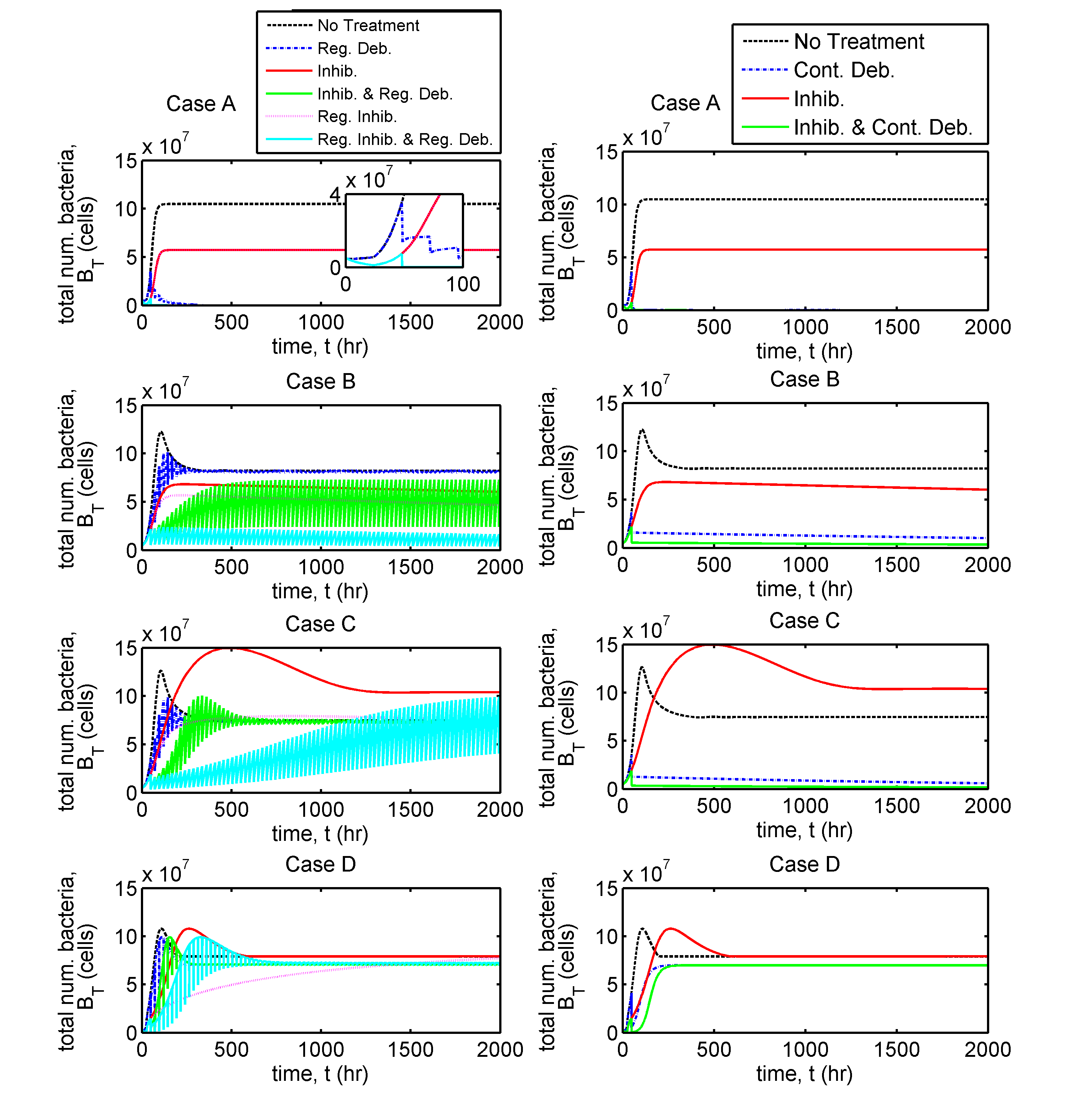}
\end{center}
\caption{Comparison of bacterial population dynamics under a variety of treatment regimes. The total number of bacteria, $B_T(t)$ ($= VB_F(t) + AB_B(t)$), is plotted in each case. Simulations extend beyond the time span of the experiments, to 2000 hours $\approx$ 83 days. The untreated and single inhibitor dose (`Inhib.') scenarios are identical to those in Fig.\ \ref{Fig_NT_and_Treat}. See Section \ref{SubSec_Treat_Strat} for a description of each treatment strategy. Case A: all treatments except single and regular inhibitor (`Reg. Inhib.') doses eradicate the bacterial population (such that $B_T(t) < 1$); Case B: regular inhibitor doses with regular debridement (`Reg. Inhib. and Reg. Deb.') and a single inhibitor dose with continuous debridement (`Inhib. and Cont. Deb.') are most effective, reducing the bacterial population size by an order of magnitude or more; Case C: only treatments involving continuous debridement (`Cont. Deb.' and `Inhib. and Cont. Deb.') are effective, reducing the bacterial population size by an order of magnitude or more; Case D: no treatment strategy is effective. Eqs.\ \eqref{Eqn_1}--\eqref{Eqn_7} were solved using \textsf{ode15s}. See Tables \ref{Table_Param_1} and \ref{Table_Param_2} for parameter values.}
\label{Fig_Treat_Strat}
\end{figure}

In each case we present results to show the evolution in the total number of bacteria, $B_T(t) = VB_F(t) + AB_B(t)$ (Figs.\ \ref{Fig_NT_and_Treat} and S4), the total numbers of free and bound bacteria, $\hat{B}_F(t) = VB_F(t)$ and $\hat{B}_B(t) = AB_B(t)$, free and bound inhibitors, $\hat{A}_F(t) = VA_F(t)$ and $\hat{A}_B(t) = AA_B(t)$, and free binding sites, $\hat{E}(t) = AE(t)$ (Figs.\ \ref{Fig_Dep_Var}, S5 and S6), and of the individual terms in Eqs.\ \eqref{Eqn_1}--\eqref{Eqn_4} (Figs.\ S7--S12) in the untreated and single inhibitor dose scenarios. We also present results to show the evolution in the total number of bacteria in the treatment scenarios involving regular inhibitor doses and regular or continuous debridement (Figs.\ \ref{Fig_Treat_Strat}, S13 and S14). Lastly, we present a sensitivity analysis showing the effects of a tenfold increase or decrease in each of the 13 fitted parameters (Figs.\ S15-S22, see Section \ref{SubSec_SA} for details).

In the remainder of this paper, we distinguish between rate constants, e\@.g\@., $\alpha_{Bac}$ and $\delta_B$, and the rate at which processes occur, e\@.g\@., $\alpha_{Bac}AB_FE$ and $\delta_BB_B$, the former being distinguished from the latter by the use of the word `constant'. We also distinguish between the intrinsic growth rate, e\@.g\@., $r_F$, and the rate of logistic growth, e\@.g\@., $r_FB_F(1-B_F/K_F)$.
\subsubsection{Case A}\label{SubSubSec_Case_A}
\noindent
Distinguishing features of the model predictions:
\begin{itemize}
\item The total numbers of bacteria in both the untreated and single inhibitor dose scenarios are essentially monotone increasing functions of time, achieving their steady-state values earlier than in the other cases;
\item All treatments are effective in reducing the number of bacteria;
\item All treatments except for single and regular inhibitor doses eliminate the bacterial population (such that $B_T(t) < 1$) by $t = 2000$ hr.
\end{itemize}
Explanation for the effect of treatment with inhibitors:
\\\\
Case A includes parameter sets 1 and 2 (see Supplementary Material, Section S2), where the results from Set 2 are presented in the main text. The evolution in the total number of bacteria, $B_T(t)$, is qualitatively similar for both sets in the untreated scenario and similarly in the single inhibitor dose scenario (see Figs.\ \ref{Fig_NT_and_Treat} and S4); however, interestingly, the parameter sets achieve this total by different means in the untreated scenario. In Set 1, the majority of bacteria are predicted to be free in the absence of treatment, while in Set 2, both free and bound bacteria are present in similar numbers (see Figs.\ \ref{Fig_Dep_Var} and S5). Thus, Set 2 is more biologically realistic than Set 1, since we would expect a significant number of bacteria (at least $O(E_{init}\times10^{-1})$) to bind to the host in the absence of inhibitor.

Surprisingly, the total number of free bacteria, $\hat{B}_F(t)$, is predicted to exceed its carrying capacity, $VK_F$, at steady-state, in the untreated scenario, in both Sets 1 and 2 (and thus the logistic growth term becomes a death term). This is valid since, as noted in Section \ref{Sec_Mod_Form}, the carrying capacities in this model represent the maximum number of bacteria that can be supported with nutrients in each compartment, rather than the maximum number that can fit into each compartment. Examination of the sizes of the terms in Eq.\ \eqref{Eqn_1} (see Fig.\ S7) reveals that this is due to the contribution of daughter cells from the bound compartment, the bacterial binding and unbinding rates essentially balancing each other. We note that the maximum proportion of bound daughter cells to enter the bound compartment, $\eta_{max}$, ranges between $O(10^{-10})$ and $O(10^{-2})$, across the 12 parameter sets considered (see Tables \ref{Table_Param_1} and S1). Therefore, the majority of bound daughter cells enter the exudate in all cases (though, once there, they will not continue to divide if the free carrying capacity is exceeded). This insight is not intuitively obvious, demonstrating the benefit of mathematical modelling.

In Case A, inhibitors are predicted to bind rapidly to the surface, achieving quasi-steady-state in 15--30 mins (see Figs.\ \ref{Fig_Dep_Var} and S6). This reduces the number of bound bacteria, compared with the untreated scenario, and, as a consequence, reduces the contribution of daughter cells to the exudate from the bound compartment (compare Figs.\ S7 and S9). In Set 2, the treatment-induced reduction in $B_T(t)$ is due largely to the drop in the number of bound bacteria as a result of replacement by inhibitors (the drop in $B_F(t)$ being small), while in Set 1 the latter effect is the more significant, the reduced flux of bound daughter cells into the free compartment resulting in a significant reduction in the number of free bacteria.

Sensitivity analysis shows that $B_T(t)$ is most sensitive to the free carrying capacity, $K_F$, in both the untreated and single inhibitor dose scenarios, showing no significant sensitivity to any other fitted parameter in the single inhibitor dose scenario (see Figs.\ S15 and S16). This parameter is significant in the latter scenario since the majority of bacteria are in the free compartment when treatment is applied, achieving their carrying capacities. This also suggests that regular and continuous debridement should both be effective in reducing the bacterial population size.
\\\\
Exploring other treatment strategies:
\\\\
Regular inhibitor dosing results in a total bacterial population size almost identical to that in the single inhibitor dose case. Each of the other treatment regimes examined are predicted to be successful in eliminating the bacterial population (such that $B_T(t) < 1$) given sufficient time --- regular debridement takes the longest at 200 hr (Set 1) or 1700 hr (Set 2), while continuous debridement takes 70 hr (Set 1) or 140 hr (Set 2) and treatments combining inhibitor with regular or continuous debridement are much more rapid, elimination of bacteria occurring upon the initial debridement event at 48 hr (see Figs.\ \ref{Fig_Treat_Strat}, S13 and S14).

In both Sets 1 and 2, the rate constant of bacterial unbinding is high (see Tables \ref{Table_Param_1} and S1), so that even when inhibitors are not used, regular and continuous debridement are effective in eliminating bacteria. Sensitivity analysis of treatment strategies 3--8 shows that regular inhibitor dosing is sensitive to $K_F$, similar to treatment with a single inhibitor dose, while regular debridement is sensitive to $r_F$, $r_B$, $\alpha_{Bac}$, $\beta_{Bac}$ and $\tilde{\psi}_{Bac}$, parameter variation resulting in an increase in $B_T(t)$ in all those cases where the change is significant. The remaining treatment strategies are almost entirely insensitive to variation in the fitted parameters (Figs.\ S17--S22) and hence relatively robust.
\subsubsection{Case B}\label{SubSubSec_Case_B}
\noindent
Distinguishing features of the model predictions:
\begin{itemize}
\item Single and repeated inhibitor doses, continuous debridement and a single inhibitor dose with continuous debridement all result in a long-term reduction in the total number of bacteria;
\item The use of regular inhibitor doses with regular debridement is consistently effective in significantly reducing (parameter sets 3, 4, 6 and 7) or removing (parameter set 5) the bacterial population;
\item Regular debridement and a single inhibitor dose with regular debridement are not consistently effective in reducing the bacterial burden.
\end{itemize}
Explanation for the effect of treatment with inhibitors:
\\\\
Case B includes parameter sets 3--7 (see Supplementary Material, Section S2), where the results from Set 3 are presented in the main text. The evolution in the total number of bacteria, $B_T(t)$, is qualitatively similar for each set in the untreated scenario and similarly in the single inhibitor dose scenario (see Figs.\ \ref{Fig_NT_and_Treat} and S4). In the untreated scenario, $B_T(t)$ achieves an early maximum at about 100 hrs, before settling to a reduced steady-state. The same behaviour is present in Cases C and D (see Figs.\ \ref{Fig_NT_and_Treat} and S4). It can be seen from Figs.\ \ref{Fig_Dep_Var} and S5 that the drop in $B_T(t)$ is caused by a drop in $B_F(t)$, the bound bacterial population size being essentially monotone increasing. Examination of the terms in Eq.\ \eqref{Eqn_1} in Cases B--D reveals that the early growth in free bacterial numbers is due to the contribution of daughter cells from the bound compartment, while the subsequent reduction in numbers occurs as the bound bacteria reach or exceed their carrying capacity, $K_B$ (see Fig.\ S7). At this point the supply of bacteria to the free compartment from the bound compartment is greatly diminished or ceases, and the number of free bacteria drops through binding to the surface and (with the exception of Case C, Set 10) death, free bacteria having exceeded their carrying capacity, $K_F$.

\begin{figure}
\begin{center}
\includegraphics[scale=0.175]{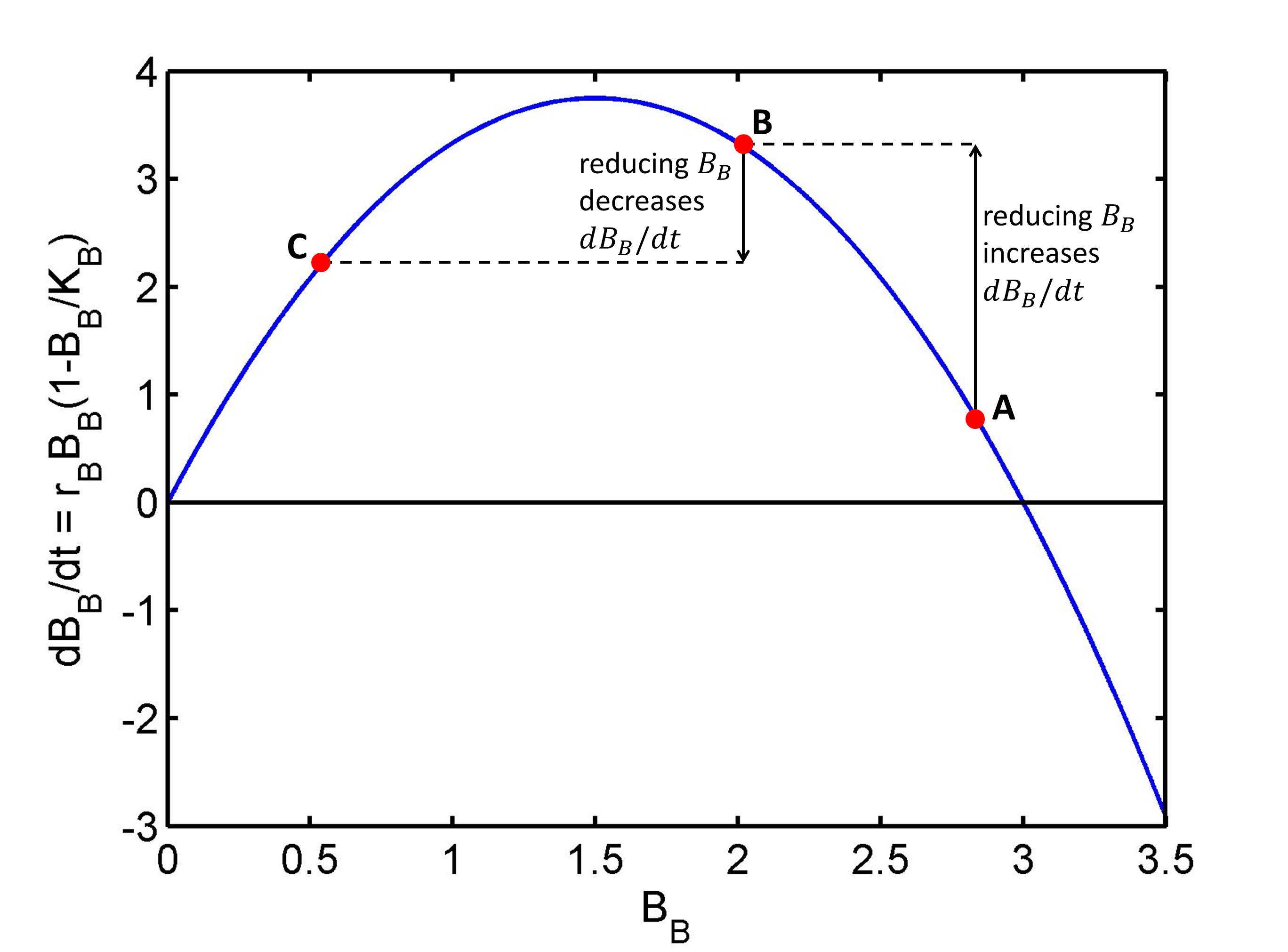}
\end{center}
\caption{Diagram demonstrating the effect of a drop in the density of bound bacteria, $B_B(t)$, upon the logistic growth rate of bound bacteria, $\mathrm{d}B_B/\mathrm{d}t = r_BB_B(1 - B_B/K_B)$. Treatment with inhibitor results in a drop in $B_B(t)$. The effect of this reduction in $B_B(t)$ upon the logistic growth rate depends upon the present value of $B_B(t)$, the value to which it is reduced and the carrying capacity, $K_B$. In this example, a reduction in $B_B(t)$ from A to B results in an increase in the logistic growth rate, while a reduction from B to C results in a decrease in the logistic growth rate. Parameter choices are for demonstration purposes only.}
\label{Fig_Log_Growth}
\end{figure}

Bound bacteria are predicted to be present in significant numbers for all parameter sets in Case B, outnumbering free bacteria at steady-state in parameter sets 3--5 (see Figs.\ \ref{Fig_Dep_Var} and S5). Treatment with inhibitors results in a significant reduction in the steady-state population size of bound bacteria (see Figs.\ \ref{Fig_Dep_Var} and S6). However, surprisingly, it may either increase (Sets 3, 4 and 6) or decrease (Sets 5 and 7) the number of free bacteria, depending upon the effect of treatment upon the logistic growth of bound bacteria. Due to the quadratic dependence of the logistic growth rate of bound bacteria upon their density, a drop in the density of bound bacteria may either increase of decrease the rate of logistic growth and hence the flux of bound daughter cells into the free compartment. This is determined by the densities between which the bound compartment shifts and the bound carrying capacity (see Fig.\ \ref{Fig_Log_Growth}). Despite this, a single inhibitor dose is effective in reducing the total bacterial population size in all parameter sets, since the loss in bound bacteria is greater than the gain in free bacteria.

We note that the inhibitor binding rate constants in Case B are 4--5 orders of magnitude lower than those in Case A (see Tables \ref{Table_Param_1} and S1) and consequently inhibitors take longer to reach quasi-steady-state --- on the order of 100--200 hrs compared with 15--30 mins for Case A. Having reached quasi-steady-state, inhibitors continue to bind at a lower rate than during the initial rapid phase (see Figs.\ \ref{Fig_Dep_Var} and S6, and see Cases C and D below for comparison).

Sensitivity analysis of the single inhibitor dose scenario shows that $B_T(t)$ is most sensitive to $K_F$, with significant sensitivity to $\alpha_{Bac}$, $\delta_B$ and $\alpha_A$ in Sets 3 and 4 (see Fig.\ S16). The free carrying capacity is significant since most bacteria are in the free compartment when treatment is applied. By contrast, $B_T(t)$ is most sensitive to $K_B$ in the untreated scenario (see Fig.\ S15).
\\\\
Exploring other treatment strategies:
\\\\
Regular debridement alone is predicted to be ineffective since there are a significant number of bound bacteria at the untreated steady-state and the rate constants of bacterial unbinding are much lower than in Case A (see Tables \ref{Table_Param_1} and S1); however, continuous debridement is effective since free bacteria are eliminated while bound bacteria unbind gradually from the surface and are cleared (see Figs.\ \ref{Fig_Treat_Strat}, S13 and S14). Combining continuous debridement with inhibitors improves efficacy, compared with continuous debridement alone, as would be expected (see Fig.\ S14). Of those treatments which involve regular debridement, that which includes regular inhibitor doses is most effective, the inhibitors serving to displace the bound bacteria and debridement to remove free bacteria.

Combining inhibitor with continuous debridement is more effective than inhibitor alone, in all cases except Set 7, where the addition of continuous debridement results in an increased $B_T(t)$ (see Fig.\ S14). This increase may be explained by the significant contribution made to $B_B(t)$ by logistic growth in Set 7, in the single dose scenario, whereas binding plays a more significant role in Sets 3--6 (see Fig.\ S10). By removing free inhibitors before they have finished binding, more bound daughter cells may colonize the surface, increasing the bound population size, and hence $B_T(t)$, above that in the single inhibitor dose scenario.

The regular inhibitor dosing scenario is more sensitive to changes in the parameters than the single inhibitor scenario, though it does not show a consistently strong sensitivity to any one parameter (Fig.\ S17). The regular debridement scenario is most sensitive to $K_B$ (Fig.\ S18), consistent with the observation that regular debridement is more effective when combined with inhibitor, which may have a similar effect to reducing $K_B$, while treatments combining inhibitor with regular debridement are most sensitive to $r_B$ in general (Figs.\ S19 and S20). Treatments involving continuous debridement are relatively insensitive to parameter changes for the most part, showing no consistent sensitivity to any one parameter (Figs.\ S21 and S22).
\subsubsection{Case C}\label{SubSubSec_Case_C}
\noindent
Distinguishing features of the model predictions:
\begin{itemize}
\item Single, and in some cases (parameter sets 8 and 10) repeated, inhibitor doses increase the total bacterial population size.
\end{itemize}
Explanation for the effect of treatment with inhibitors:
\\\\
Case C includes parameter sets 8--11 (see Supplementary Material, Section S2), where the results from Set 8 are presented in the main text. The evolution in the total number of bacteria, $B_T(t)$, is qualitatively similar for each set in the untreated scenario and similarly in the single inhibitor dose scenario (see Figs.\ \ref{Fig_NT_and_Treat} and S4). In the untreated scenario, $B_T(t)$ evolves in a qualitatively similar way to Cases B and D; however, we have the surprising result that application of a single inhibitor dose leads to an increase in $B_T(t)$ at steady-state.

This last result may be explained as follows. The number of bound bacteria exceeds that of free bacteria in the untreated scenario for all parameter sets in Case C (see Figs.\ \ref{Fig_Dep_Var} and S5). Application of a single inhibitor dose results in a slight reduction in the number of bound bacteria in Sets 8, 9 and 11 and a slight increase in Set 10, while the number of free bacteria increases by an order of magnitude or more in all sets (Fig.\ S6). Inhibitors bind more rapidly to the epithelium than bacteria, with $A_B(t)$ achieving its maximum value between 40--120 hr. However, whereas $A_B(t)$ is a monotone increasing function in Cases A and B, in Case C it slowly decreases, subsequent to the initial, more rapid, binding phase. Inhibitors outcompete bacteria for binding sites initially because they greatly outnumber them; however, as the number of free bacteria increases, bacteria begin to replace inhibitors. This is because, while the binding rate constants of bacteria and inhibitors are of the same order of magnitude, inhibitors have much higher unbinding rate constants, such that their association constant is much smaller (see Tables \ref{Table_Param_1} and S1). The difference between association constants is less significant in Set 11, hence the effect is less pronounced in this case (see Fig.\ S6).

Bacteria are predicted to populate the bound compartment mainly through binding (as opposed to logistic growth) in both the untreated and single inhibitor dose scenarios, but the binding rate is lower in the treated scenario, since inhibitors reduce the number of free binding sites (see Figs.\ S8 and S10). Consequently, it takes bound bacteria longer to reach their steady-state value, during which time the logistic growth of bound bacteria is maintained at a higher level (the population size being kept well below carrying capacity at which logistic growth would go to zero), the majority of daughter cells entering the free compartment, pushing the population size of free bacteria well above untreated levels. As bound bacteria approach their carrying capacity their logistic growth rate diminishes, reducing the supply of bacteria to the free compartment and the system settles to steady-state.

Set 10 is peculiar in that a single inhibitor dose increases the numbers of both bound and free bacteria, despite the fact that bound bacteria exceed their carrying capacity in the absence of treatment (compare Figs.\ S5 and S6 and see also Fig.\ S3). Treatment causes an increase in the number of free bacteria by reducing the number of free binding sites, which reduces the rate at which free bacteria bind to the surface, this being the main sink term in the untreated scenario (see Figs.\ S7 and S9). As the solution approaches steady-state, the rate of binding of free bacteria comes to exceed that in the untreated scenario, despite the decreased number of binding sites, since the number of free bacteria is now almost three orders of magnitude larger than in the untreated scenario. As a consequence of the increased rate of binding of free bacteria, the steady-state number of bound bacteria exceeds that in the untreated scenario.

Sensitivity analysis shows that the system is most sensitive to $K_B$ in the untreated scenario (Fig.\ S15), while in the single inhibitor dose scenario, parameter sets show significant sensitivity to $r_F$, $r_B$, $K_F$ and $K_B$ (Fig.\ S16). This suggests that, in Case C, treatment efficacy will depend to a large extend upon the bacterial population size that can be supported by the bound and free compartments and the rate at which the bacterial species can reproduce or die. Counter-intuitively, an increase in $r_F$ (the intrinsic growth rate of free bacteria) typically results in a decrease in $B_T(t)$. This is because the logistic growth term acts mainly as a death term, removing free bacteria from the exudate as their population size is pushed above its carrying capacity by the flux of daughter cells from the bound compartment.
\\\\
Exploring other treatment strategies:
\\\\
Regular debridement and/or inhibitor doses are ineffective in significantly reducing the bacterial population size, except in Set 11, where regular inhibitor doses with regular debridement reduced the population size by three orders of magnitude (Figs.\ \ref{Fig_Treat_Strat} and S13). Continuous debridement with and without inhibitor is effective in all cases except Set 10 where $B_T(t)$ tends to untreated levels at steady-state (Figs.\ \ref{Fig_Treat_Strat} and S14). Treatments involving continuous debridement reduce $B_T(t)$ by an order of magnitude in Sets 8 and 9 and to $O(100)$ or below in Set 11. No treatment is successful in completely eliminating the bacteria burden.

Regular inhibitor dosing is most sensitive to $r_F$, $r_B$, $K_F$, $K_B$, $\alpha_{Bac}$, $\delta_B$, $\alpha_A$ and $\beta_A$ (Fig.\ S17). Treatments involving regular debridement are generally most sensitive to $K_B$ (Figs.\ S18--S20), while those involving continuous debridement are not consistently sensitive to any one parameter (Figs.\ S21 and S22).
\subsubsection{Case D}\label{SubSubSec_Case_D}
\noindent
Distinguishing features of the model predictions:
\begin{itemize}
\item A single inhibitor dose causes the total bacterial population size to temporarily exceed that without treatment and is close to that without treatment at steady-state, though it matches the experimental result over the first six days;
\item All treatments are ineffective, the long-term bacterial population size being close to that without treatment in all cases.
\end{itemize}
Explanation for the effect of treatment with inhibitors:
\\\\
Case D consists of parameter set 12, which is presented in the main text. In the untreated scenario, $B_T(t)$ evolves in a qualitatively similar way to Cases B and C, while application of a single inhibitor dose has little effect on the steady-state value of $B_T(t)$ (see Figs.\ \ref{Fig_NT_and_Treat} and S4). The number of bound bacteria exceeds that of free bacteria in the untreated scenario (see Figs.\ \ref{Fig_Dep_Var} and S5). Application of a single inhibitor dose slows the dynamics of the system, but results in essentially the same numbers of free and bound bacteria at steady-state (see Figs.\ \ref{Fig_Dep_Var}, S5 and S6). In both the untreated and treated scenarios the growth in the population size of free bacteria is due to the flux of bound daughter cells as in Cases B and C (see Figs.\ S7 and S9). However, unlike in Cases A--C (with the exception of  Set 7), the increase in the number of bound bacteria is primarily due to their logistic growth, rather than the binding of free bacteria (see Figs.\ S8 and S10). Inhibitors bind rapidly initially, reaching quasi-steady-state in a couple of hours. This is followed by a more gradual period of unbinding as bacteria outcompete inhibitors for binding sites (see Figs.\ \ref{Fig_Dep_Var} and S6). Inhibitors slow the dynamics by reducing the rate at which bacteria bind to the surface and by reducing the proportion of bound daughter cells which can remain bound to the surface.

Sensitivity analysis shows that the system is most sensitive to $K_F$ and $K_B$ in both the untreated and single inhibitor dose scenarios (see Figs.\ S15 and S16). This is intuitive, given that $B_F(t)$ and $B_B(t)$ take values close to carrying capacity at steady-state in both scenarios.
\\\\
Exploring other treatment strategies:
\\\\
Treatment involving regular inhibitor doses, regular debridement or continuous debridement are all ineffective in Case D (see Figs.\ \ref{Fig_Treat_Strat}, S13 and S14). Regular inhibitor dosing shows greatest sensitivity to $r_F$ and $K_F$, while treatments involving regular or continuous debridement show significant sensitivity to $K_B$ only (Figs.\ S17--S22). All treatments would be improved if $K_B$ could be reduced in some way.
\subsubsection{Inhibitor sensitivity analysis}\label{SubSubSec_2DSA}
We have seen above that treatment is always effective in Case A, is sometimes effective in Cases B and C and is never effective in Case D. The question then arises: can we modify the treatment strategies in such a way as to make at least some of them effective in all cases? In this section we explore two ways in which this might be achieved. Firstly, we investigate how treatment efficacy varies with changes in the binding and unbinding rate constants of inhibitors ($\alpha_A$ and $\beta_A$); and, secondly, we examine the effect of increasing the number of inhibitors used in each inhibitor dose (see Section \ref{SubSec_SA} for details). Since we are varying parameters which relate only to the inhibitors, we consider only those five treatment scenarios that involve dosing with inhibitor.

\begin{figure}
\begin{center}
\includegraphics[scale=0.6]{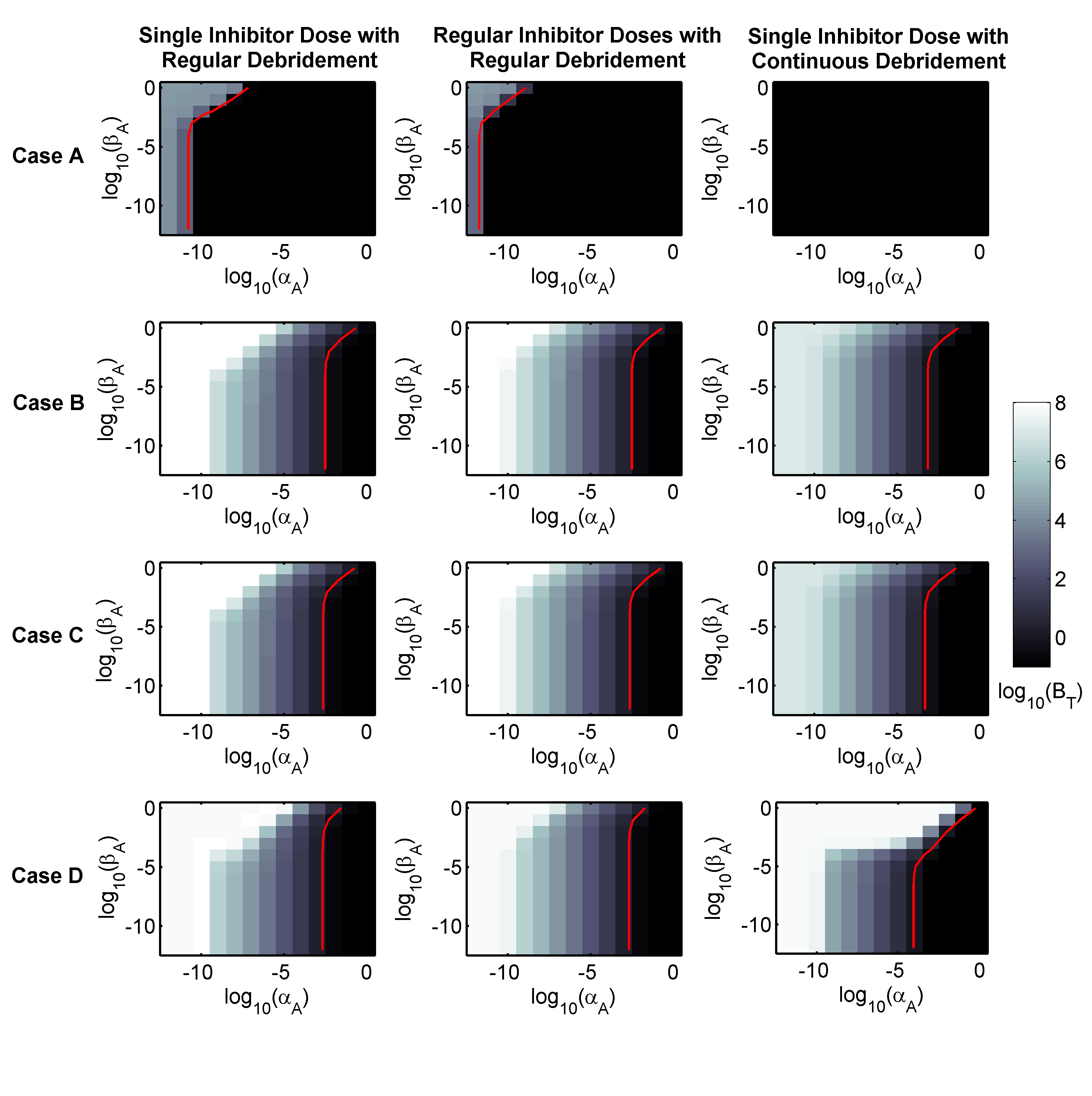}
\end{center}
\caption{Sensitivity analysis to determine the effect of varying inhibitor binding ($\alpha_A$) and unbinding ($\beta_A$) rate constants upon the efficacy of single inhibitor dose with regular debridement, regular inhibitor doses with regular debridement and single inhibitor dose with continuous debridement treatments in Cases A--D, where $A_{F_{init}} = 6.12 \times 10^7$ inhibitors cm$^{-3}$ (the standard value). For each panel, $\alpha_A$ and $\beta_A = 10^{-12}$, $10^{-11}$, \ldots, $10^{-1}$ and 1, where a $\log_{10}$ scale is used on both axes. The value of $\log_{10}(B_T(672))$ is plotted at each point in parameter space, where $B_T(672)$ ($= VB_F(672) + AB_B(672)$) is the total number of bacteria after 4 weeks (672 hr). The colour scheme is calibrated to maximise clarity, such that values of $\log_{10}(B_T(672))\leq-1$ appear in black. The red curve (colour online) traces the contour along which $B_T(672) = 1$, such that $B_T(672) > 1$ to the left and $B_T(672) < 1$ to the right of this curve. Treatment efficacy improves with increasing $\alpha_A$ and decreasing $\beta_A$. On average, treatment with a single inhibitor dose with continuous debridement is the most effective treatment, followed by regular inhibitor doses with regular debridement; a single inhibitor dose with regular debridement being the least effective of the three . Eqs.\ \eqref{Eqn_1}--\eqref{Eqn_7} were solved using \textsf{ode15s}. See Tables \ref{Table_Param_1} and \ref{Table_Param_2} for the remaining parameter values.}
\label{Fig_2D_SA}
\end{figure}

The panels in Figs.\ \ref{Fig_2D_SA} and S23--S32 show $\log_{10}(B_T(672))$ at each point in ($\log_{10}(\alpha_A)$,$\log_{10}(\beta_A)$) parameter space (where 672 hr $= 4$ weeks). Red curves mark the contours along which $B_T(672) = 1$ and thus demarcate the boundary between the region of parameter space to the left of the curve, where $B_T(672) > 1$ and treatment has failed to eradicate the bacterial population, and the region to the right of the curve, where $B_T(672) < 1$ and treatment has successfully eliminated the bacterial population. In this section, we consider treatment to be effective only if it eliminates the bacterial burden, such that $B_T(672) < 1$.

Examination of Figs.\ S23--S26 reveals that neither single nor regular inhibitor doses are predicted to be capable of eliminating the bacterial population for any of the 12 parameter sets considered. In contrast, the remaining three treatments (a single inhibitor dose with regular debridement, regular inhibitor doses with regular debridement and a single inhibitor dose with continuous debridement) are all predicted to be capable of eradicating the bacterial population, provided that the inhibitor binding rate constant is high enough and the unbinding rate constant is low enough. Therefore, in what follows, we focus upon these three treatment scenarios (Figs.\ \ref{Fig_2D_SA} and S27--S32).

Treatment is predicted to be successful throughout the region of parameter space considered for all three treatment scenarios in Set 1 and for the single inhibitor dose with continuous debridement scenario in Set 2 (see Figs.\ \ref{Fig_2D_SA} and S27--S32). Indeed, continuous debridement alone is successful in both Sets 1 and 2 as is regular debridement alone in Set 1 (see Figs.\ \ref{Fig_Treat_Strat}, S13 and S14). Each treatment strategy is more effective, that is bacteria are eliminated in a larger region of ($\alpha_A$,$\beta_A$) parameter space, when more inhibitors are used per dose. On average, across the 12 parameter sets, a single inhibitor dose with continuous debridement is predicted to be the most effective strategy, followed by regular inhibitor doses with regular debridement and then single inhibitor dose with regular debridement (see Figs.\ \ref{Fig_2D_SA} and S27--S32).

Tables S3 and S4 show the minimum value of $\alpha_A$, the maximum value of $\beta_A$, the minimum value of $\alpha_A/\beta_A$ and the maximum ratio of bacterial to inhibitor association constants, $(\alpha_{Bac}/\beta_{Bac})/(\alpha_A/\beta_A)$, predicted to be required for each treatment to eliminate the bacterial population for each parameter set, for both standard and higher concentration inhibitor doses. It can be seen that $\alpha_A$ must be at least $10^{-2}$ hr$^{-1}$ sites$^{-1}$ in order for each of the treatments to eliminate the bacterial burden for all parameter sets when standard inhibitor doses are used, and that a value of $10^{-3}$ hr$^{-1}$ sites$^{-1}$ will suffice where higher concentration doses are used. Treatment efficacy is less dependent upon $\beta_A$, such that even for values as large as 1 hr$^{-1}$, a value of $\alpha_A$ can be found such that treatment eliminates the bacterial population for all treatment strategies, doses and parameter sets.

The model predicts that the inhibitor association constant, $\alpha_A/\beta_A$, must take a minimum value of 1 sites$^{-1}$ in order for treatment to eliminate the bacterial burden across all parameter sets for each of the treatments in the standard dose scenario and for single inhibitor with continuous debridement in the higher concentration dose scenario, while a value of $10^{-1}$ sites$^{-1}$ suffices for single and regular inhibitor with regular debridement in the higher concentration dose scenario. Lastly, in order for treatment to eliminate the bacterial population, the ratio of bacterial to inhibitor association constants must be no more than $\approx 10^{-7}$ in the single and regular inhibitor with regular debridement scenarios with standard doses, $\approx 10^{-6}$ in the single and regular inhibitor with regular debridement scenarios with higher concentration doses, and $\approx 6 \times 10^{-6}$ in the single inhibitor with continuous debridement scenario with standard or higher concentration doses. Thus, the inhibitor association constant must be 6--7 orders of magnitude larger than that of the bacteria in order for treatment to eliminate the bacterial population in all cases, though the association constant can be much lower in many parameter sets (e\@.g.\ in Sets 5, 9 and 10, the inhibitor association constant can be $O(10)$--$O(100)$ times smaller than the bacteria association constant for all three treatments in the higher concentration dose scenario).

\begin{figure}
\begin{center}
\includegraphics[scale=0.59]{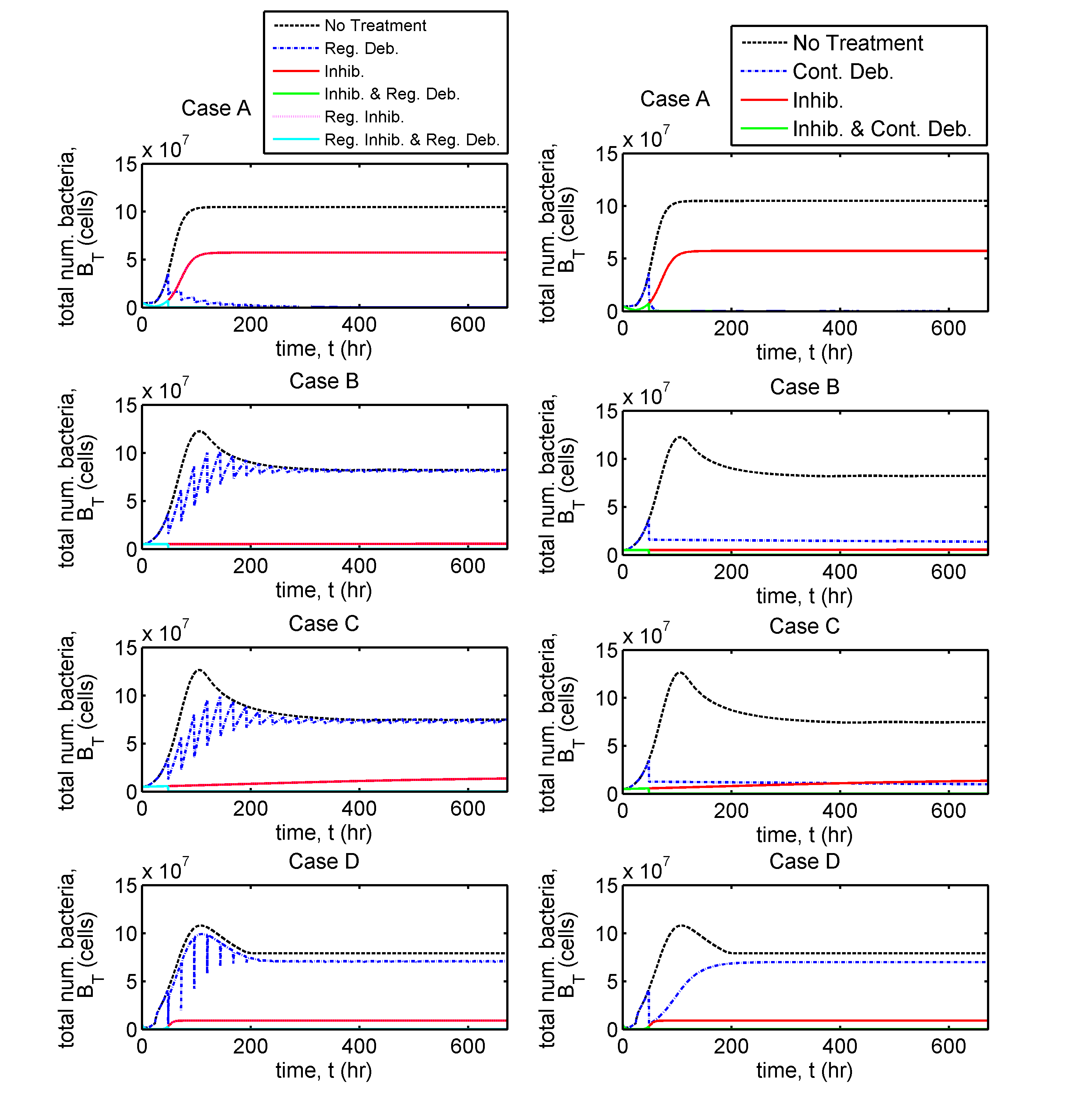}
\end{center}
\caption{Comparison of bacterial population dynamics under a variety of treatment regimes with modified inhibitors. The total number of bacteria, $B_T(t)$ ($= VB_F(t) + AB_B(t)$), is plotted in each case. Simulations extend beyond the time span of the experiments, to 672 hours = 28 days. Bacteria are eliminated ($B_T(t)<1$) by $t = 672$ hr for all treatment strategies combining inhibitor with regular or continuous debridement in all cases. In Case D, $B_T(672)>1$; however, it descends below 1 at an earlier time point so bacteria may be considered to have been eliminated by $t = 672$. Eqs.\ \eqref{Eqn_1}--\eqref{Eqn_7} were solved using \textsf{ode15s}. Parameter values: $\alpha_A = 10^{-2}$ and $\beta_A = 10^{-2}$. See Tables \ref{Table_Param_1} and \ref{Table_Param_2} for the remaining parameter values.}
\label{Fig_Treat_Strat_Mod_Bead}
\end{figure}

Fig.\ \ref{Fig_Treat_Strat_Mod_Bead} shows the evolution in the total number of bacteria, $B_T(t)$, for the full range of treatment strategies in Cases A--D, where $\alpha_A = 10^{-2}$ and $\beta_A = 10^{-2}$. These values are chosen, based upon the sensitivity analysis above, such that all those treatment strategies which combine inhibitor with regular or continuous debridement are effective in eliminating the bacterial population, that is $B_T(672)<1$. The only exception is for the single inhibitor dose with continuous debridement treatment in Case D, where $B_T(672)\approx 1.07$; however, $B_T(t)<1$ for $t\in(48,665)$, so treatment can be considered to have eliminated the bacterial population in this case also. Therefore, our models make the encouraging prediction that, in theory, treatment could be rendered successful in all cases.
\section{Discussion}\label{Sec_Disc}
As bacteria gain increasing resistance to antibiotics it is vital that we develop alternative treatment strategies. Anti-virulence treatments --- specifically MAM7-coupled beads, which operate by competitively inhibiting the binding of bacteria to host cells --- present a promising complement or alternative to antibiotics. Opinion as to the likely efficacy of such treatments is mixed, with some suggesting that their utility may be limited to preventing the initiation of a bacterial infection (prophylaxis) as opposed to treating a pre-existing infection (therapy) \citep{Cegelski_et_al_2008,Clatworthy_et_al_2007,Krachler_and_Orth_2013,Zambelloni_et_al_2015}. In this paper we have used mathematical models to help us interpret the results of an experimental model, involving the inhibitor treatment of a burn wound infected by \emph{P.\ aeruginosa} in the rat (see Section \ref{Sec_Exp_Set}). Our models allow us to predict the conditions under which treatment with inhibitors will be effective and to explore ways in which inhibitor dosing could be augmented to improve efficacy.

Mathematical models were fitted to experimental data using a combination of MCMC and frequentist techniques (Section \ref{SubSec_Para_Fit}). The data was insufficiently detailed to obtain a unique model fit; however, a number of close fits were obtained (12 parameter sets are explored here) and classified into four qualitatively different cases (A--D).

Given the significant qualitative, and not merely quantitative, differences in predicted treatment outcomes between Cases A--D, this work highlights the importance of considering a range of viable parameter sets. Had a single parameter set been chosen, our conclusions would have differed markedly from those in this more comprehensive study. It may be that different parameters sets could reflect inter-patient variability, as well as different bacterial species. Indeed, a variety of Gram negative and Gram positive bacteria have been found to infect burn wounds \citep[see, for example,][]{Azzopardi_et_al_2014,Church_et_al_2006,Oncul_et_al_2014,Weber_et_al_1997}.

Eight treatment scenarios were considered for each of Cases A--D (see Section \ref{SubSec_Treat_Strat} for details). The untreated and single inhibitor dose scenarios were considered in the experimental model, while the rest are theoretical treatments that remain to be tested experimentally. Continuous debridement, with clearance rates on the order of magnitude of those used in simulations, is probably not practically achievable; though it might be possible to maintain a high level of clearance using negative pressure wound therapy (see Section \ref{Sec_Intro}). In any case, these simulations allow us to determine the theoretical best-case scenario where debridement is applied.

Steady-state analysis of the untreated and single inhibitor dose scenarios reveals that they each have a single stable steady-state to which the system will proceed (Section \ref{SubSec_StStAnal}). The stable steady-states take the form of either stable (improper) nodes or stable spirals, suggesting a means of empirically distinguishing between parameter sets were the oscillations experimentally detectable. The monostability of the system also suggests that inhibitor treatment would be effective when used therapeutically (not just prophylactically) since the system must shift to its new stable steady-state upon treatment (though there will be a delay in reaching it).

Simulations for the 8 treatment scenarios in Cases A--D reveal a range of outcomes and  provide insight into the bacterial population dynamics (Sections \ref{SubSubSec_Case_A}--\ref{SubSubSec_Case_D}). Before considering each case in turn, a few general observations can be made. Firstly, while the ratio between the number of bacteria at carrying capacity in the free and bound compartments, $VK_F$ and $AK_B$ respectively, varies between parameter sets, the intrinsic growth rate in the bound compartment, $r_B$, is consistently greater than that of the free compartment, $r_F$ (see Tables \ref{Table_Param_1} and S1). This suggests that the bound compartment is more favourable to the growth of bacteria than the free compartment, agreeing nicely with \citeauthor{Huebinger_et_al_2016}'s speculative explanation for the efficacy of inhibitor treatment. This might be because bound bacteria have access to additional nutrients derived from the epithelial cells, not available to free bacteria.

The proportion of bound daughter cells to enter the bound compartment, $\eta_{max}$, takes values in the range $[10^{-10},10^{-2}]$ across the 12 parameter sets (see Tables \ref{Table_Param_1} and S1). As such, the majority of the bound daughter cells enter the free compartment. As can be seen from Figs.\ S7 and S9, it is the input of bound daughter cells that makes the most significant contribution to the free compartment in most parameter sets in both the untreated and single inhibitor dose scenarios. This observation is key in understanding the effects of inhibitor treatment. If inhibitors bind sufficiently rapidly and numerously to the epithelium, then the number of bacteria in the bound compartment and hence their rate of logistic growth will remain small. As a result, the flux of bound daughter cells into the free compartment will be diminished, reducing the number of bacteria in the free compartment at steady-state. In this case, inhibitor treatment will reduce the total number of bacteria at steady-state. However, if inhibitors bind in lower numbers, then the number of bound bacteria at steady-state may be reduced, dropping it further beneath carrying capacity, but maintaining a large population size. In this case, the logistic growth of bound bacteria is maintained at a higher rate (see Figure \ref{Fig_Log_Growth}) as is the flux of bound daughter cells into the free compartment, with the result that inhibitor treatment increases the total number of bacteria at steady-state above its untreated value. Therefore, our model predicts that in order for inhibitor treatment to be effective, inhibitors must bind rapidly and numerously. If they do not, then treatment may actually worsen a bacterial infection.

If debridement is applied too early, as in Set 7, it may reduce treatment efficacy as inhibitors will have had insufficient time to bind before those in the free compartment are removed. If inhibitor binding rates could be determined experimentally then this would allow us to determine the optimum timing of debridement in relation to inhibitor dosing.

Considering Cases A--D in turn. All treatments are effective in significantly reducing the bacterial population size in Case A, with all treatments involving regular or continuous debridement eliminating the bacterial population. In Case B, all treatments except regular debridement and a single inhibitor dose with regular debridement are consistently effective; a single inhibitor dose with continuous debridement eliminating, and continuous debridement and regular inhibitor doses with regular debridement almost eliminating the bacterial population by 80 days. Counterintuitively, a single inhibitor dose increases the bacterial burden in Case C, with treatments involving continuous debridement being the most effective, though no treatment is effective in Set 10. Lastly, in Case D, all of the treatment strategies result in a steady-state bacterial population size similar to that without treatment.

The differences in behaviour between Cases A--D may be explained, to some extent, by the differences in the bacteria and inhibitor association constants and the ratio of these constants (see Tables \ref{Table_Param_1} and S1). The bacteria association constant is lower and the inhibitor association constant higher in Case A than in the other cases, so that inhibitors are much more successful in out-competing bacteria for binding sites. While the bacteria association constant is greater in Cases B--D than in A (with the exception of Set 6 which is the same order of magnitude as Set 2), the values taken in Cases B--D are overlapping. Therefore, the differences in behaviour between cases are not easily explained by this parameter alone. The inhibitor association constant is $O(10^1)$--$O(10^6)$ in Case A, $O(10^{-5})$--$O(1)$ in Case B and $O(10^{-7})$--$O(10^{-8})$ in Cases C and D, while the ratio between association constants is $O(10^{-17})$--$O(10^{-10})$ in Case A, $O(10^{-4})$--$O(1)$ in Case B and $O(1)$--$O(10^7)$ in Cases C and D. Interestingly, the ratio of association constants takes the same value to 4 significant figures in Sets 3--5. Therefore, both the inhibitor association constant and the ratio of association constants would seem to correlate with the differences in behaviour seen between Case A, Case B and Cases C and D; treatments involving inhibitor being more effective where the inhibitor association constant is higher and the bacterial association constant is lower, and hence where the ratio of association constants is lower. The qualitative difference between Cases C and D is less significant than between the other cases, thus it is not surprising that the order of magnitude of their association constants and their ratio are not distinct.

Given the correlation between bacteria and inhibitor association constants and treatment efficacy, a natural way to seek to improve treatment would be to decrease the bacterial association constant or to increase the inhibitor association constant, shifting the system behaviour towards that in Case A. The bacterial association constant could be reduced by treating with a molecule that binds to and blocks bacterial adhesins used for binding to the epithelium e\@.g.\ mannosides, which bind FimH \citep{Spaulding_et_al_2017}. In order for these molecules not to interfere with the inhibitor treatment they would need to either bind to an adhesin other than MAM7 or be applied to a wound before the inhibitor dose, giving the molecules time to bind to the bacteria. The inhibitor association constant could be increased by coupling more MAM7 molecules to each bead.

Sensitivity analysis of the inhibitor binding rate constant, unbinding rate constant and dose concentration predicts conditions under which treatments involving inhibitor would eliminate the bacterial burden within 28 days (Section \ref{SubSubSec_2DSA} and Figs.\ \ref{Fig_2D_SA} and S23--S32). Neither a single nor regular doses of inhibitor on their own were predicted to be capable of eliminating the bacterial population, though they may reduce it by several orders of magnitude. This is to be expected since, on its own, inhibitor can only prevent bacteria from entering the bound compartment, and is unable to remove bacteria from the free compartment. Encouragingly, a single inhibitor dose with regular debridement, regular inhibitor doses with regular debridement and a single inhibitor dose with continuous debridement were all predicted to have the potential to eliminate the bacterial population. Treatment is predicted to be more effective for higher binding rate constants, lower unbinding rate constants and higher concentration inhibitor doses, where treatment efficacy is more sensitive to the binding rate constant than to the unbinding rate constant. As judged by the area of $(\log_{10}(\alpha_A),\log_{10}(\beta_A))$ parameter space in which treatment eliminates the bacterial population --- larger areas corresponding to more effective treatments --- a single inhibitor dose with continuous debridement is the most effective treatment, followed by regular inhibitor doses with regular debridement, with a single inhibitor dose with regular debridement being the least effective of the three.

In future work we will proceed on two fronts: experimental and theoretical. We will conduct further experiments to test our model predictions and to improve model parametrisation. Experiments like those described in Section \ref{Sec_Exp_Set} could be performed over a longer time-span and separate measurements made for the numbers of free and bound bacteria and inhibitors. This would enable us to refine the range of possible parameter sets and perhaps to identify a unique best fit. This would also be aided by more regular wound imaging than the current daily set-up. Further experiments to test the model predictions concerning the other treatment strategies suggested here would also be valuable. We will develop our mathematical modelling in at least two directions. Firstly, we will incorporate treatment with antibiotics, seeking to determine the optimum treatment regime when combined with inhibitor dosing and regular or continuous debridement. Secondly, we will consider a discrete-stochastic cellular automata model to capture some of the mechanisms in the system in more detail, including the spatial spread of infection, which has been shown experimentally to be limited upon treatment with inhibitor \citep{Huebinger_et_al_2016}.

In conclusion, our mathematical models suggest that inhibitor treatment could be effective in eliminating or significantly reducing the bacterial burden in a burn wound when combined with regular or continuous debridement. Where inhibitor treatment is effective, it operates both by preventing bacteria from occupying the epithelium (where growth rates are predicted to be higher) and, consequently, by reducing the flux of bound daughter cells into the exudate. Our model predicts that inhibitor treatments, in particular those involving regular or continuous debridement, could be effective when used both prophylactically and therapeutically. Our models further predict that treatment efficacy can be improved by optimising inhibitor design and dosing schedules.
\section*{Acknowledgements}
PAR, EK, AMK and SJ gratefully acknowledge support from the Biotechnology and Biological Sciences Research Council (grant code: BB/M021386/1, \url{www.bbsrc.ac.uk/}). PAR and SJ would also like to thank the Wellcome Trust (grant code: 1516ISSFFEL9) for funding a parameterisation workshop at the University of Birmingham (UK). RMH acknowledges support from the Golden Charity Guild Charles R. Baxter, MD Chair in Burn Surgery (\url{www.utsouthwestern.edu}) which provided research funding. AMK thanks the UT System for support through a University of Texas System Science and Technology Acquisition and Retention (STARs) Program award (\url{www.utsystem.edu/offices/health-affairs/stars-program}).
%
\bibliography{Roberts_et_al_2017_Ref}{}
\bibliographystyle{plainnat}
\end{document}